\documentclass[journal=jpcbfk,manuscript=article,layout=twocolumn]{achemso}

\usepackage{amsfonts}
\usepackage{amssymb}
\usepackage{graphicx}
\usepackage{amsmath}
\newcommand{\beginsupplement}{%
        \setcounter{table}{0}
        \renewcommand{\thetable}{S\arabic{table}}%
        \setcounter{figure}{0}
        \renewcommand{\thefigure}{S\arabic{figure}}%
        \renewcommand\thepage{S\arabic{page}}
        \setcounter{equation}{0}
        \renewcommand{\theequation}{S\arabic{equation}}%
        \setcounter{section}{0}
        \renewcommand{\thesection}{\Alph{section}.}%
     }

\author{Lu Wang}
\affiliation[Inner Mongolia University of Science and Technology]
{School of Science, Inner Mongolia University of Science and Technology, Inner Mongolia, 014010, China}

\author{Fulu Zheng} %???
\affiliation[Max-Planck-Institut f\"ur Physik komplexer Systeme]
{Max-Planck-Institut f\"ur Physik komplexer Systeme, N\"othnitzer Strasse 38, D-01187 Dresden, Germany}

\author{Jiaming Wang}
\affiliation[Inner Mongolia University of Science and Technology]
{School of Science, Inner Mongolia University of Science and Technology, Inner Mongolia, 014010, China}
\alsoaffiliation{Division of Materials Science, Nanyang Technological University, Singapore 639798, Singapore}

\author{Frank Gro{\ss}mann}
\affiliation[Technische Universit\"at Dresden]
{Institut f\"ur Theoretische Physik, Technische Universit\"at Dresden, D-01062 Dresden, Germany}

\author{Yang Zhao}
\email{yzhao@ntu.edu.sg}
\phone{+65 6513 7990}
\fax{+65 6790 9081}
\affiliation[Nanyang Technological University]
{Division of Materials Science, Nanyang Technological University, Singapore 639798, Singapore}

\title[An \textsf{achemso} demo]
{Schr\"{o}dinger-cat states in Landau-Zener-St\"{u}ckelberg-Majorana interferometry: a multiple Davydov Ansatz approach}

\SectionNumbersOn
\begin{document}

\begin{abstract}
Employing the time-dependent variational principle combined with the multiple Davydov
$\mathrm{D}_2$ {\it Ansatz}, we investigate Landau-Zener (LZ) transitions in a qubit coupled to a photon mode with various initial photon states at zero temperature.
Thanks to the multiple Davydov trial states, exact photonic
dynamics taking place in the course of the LZ transition is also studied efficiently.
With the qubit driven by a linear external field and {the} photon mode initialized with Schr\"odinger-cat states, asymptotic behavior
of the transition probability beyond the rotating-wave approximation is uncovered for a variety of Schr\"odinger-cat initial states.
Using a sinusoidal external driving field, we also explore the photon-assisted dynamics of Landau-Zener-St\"{u}ckelberg-Majorana interferometry. Transition pathways involving multiple energy levels are unveiled by analyzing the photon dynamics.
\end{abstract}

\section{introduction}

In the course of a Landau-Zener(LZ) transition, a two-level system undergoes an avoided crossing in the presence of an external driving field \cite{Lan32, Zen32}. This celebrated physical phenomenon was independently studied by Landau, Zener, St\"uckelberg, and Majorana in 1932, and the standard LZ model is also referred to as the Landau-Zener-{St\"uckelberg}-Majorana (LZSM) model \cite{Lan32, Zen32, Stu32, Maj32}, which has found applications in a large variety of fields, including atomic and molecular physics \cite{Child,thiel_1990, lipert_1988,WeSe99,xie_2017,Gross3}, quantum optics \cite{bouw_1995}, solid state physics \cite{Wernsdorfer_2000}, chemical physics \cite{Co93,zhu_1997}, and quantum information science \cite{fuchs_2011}. Recently, as the number of physical systems described by the LZSM model grows, so does renewed attention it has received. New applications of the LZSM model have been reported in a nitrogen-vacancy center in a diamond lattice \cite{fuchs_2011}, a one-electron semiconductor double quantum dot \cite{Ota2018}, and an accelerated Bose-Einstein condensate that is synthetically spin-orbit coupled \cite{olson_2014}.

Due to its theoretical importance and wide-ranging potential applications in quantum devices, there is sustained interest in the LZSM model \cite{saito_2006,
  oliver_2005,ZHK08,niemcyzk_2010,higuchi17,militello19}. Various devices have been invented to implement a quantum system interacting with an external electromagnetic field, such
as a charge qubit coupled to a superconducting
  transmission line resonator \cite{Wallraff2004} and a superconducting flux qubit coupled to a quantum interference device \cite{chiorescu_2004}, which can be seen as
  two-level artificial atoms tuned by external fields. Such devices allow for efficient control of parameters such as the interaction strength and the bias
  \cite{Astafiev2007}. In recent years, dielectrics and semiconductors driven by strong electromagnetic field make it possible to control electron
  dynamics on the sub-femtosecond timescale \cite{higuchi17}. Electromagnetic fields can be used to detect the qubit states.
  The LZSM model is the ideal platform to depict the fundamental physics underlying these processes.

The LZSM model has also been theoretically investigated over the decades with various approaches.
For instance, a series of studies have been conducted using the LZSM model to unveil environmental effects on mechanisms of quantum electrodynamics (QED) devices.
 With the time-dependent perturbation theory,
 final transition probabilities in the fast and the slow sweeping limits have been studied by Ao and coworkers \cite{ao_1989}.
Inspired by the realization of LZSM physics in QED devices, H\"anggi and coworkers systematically investigated the final transition probabilities influenced by {a} bosonic bath at zero
temperature \cite{WSKHK06,saito_2007}. Nalbach {\it et al.}$~$extensively studied thermal effects in the dissipative LZSM model using the quasi-adiabatic
propagator path integral method and the non-equilibrium Bloch equations
\cite{nalbach_2009, nalbach_2013_prb, nalbach_2014, nalbach_2015,nalbach_2017}.
A random-variable driving approach, pioneered in Refs.~\cite{SDG99,SG02}, has been used by Stockburger \cite{St16} and Orth {\it et al.} \cite{orth_2010, orth_2013} in the present context. Furthermore, Huang and Zhao adopted the multiple Davydov trial states to elucidate the dissipative LZSM dynamics including full details on the associated boson dynamics \cite{huang2018}.

A qubit driven by an external electromagnetic field can be described by the LZSM model interacting with a single harmonic oscillator
with the frequency $\omega$ \cite{Astafiev2007, saito_2006, sun_2012, ashhab_2014, huang2018, MaRa18}.
For this model, Saito {\it et al.}$~$have revealed that zero-temperature dynamics depends strongly on the oscillator frequency only at intermediate times, if the oscillator is in
its ground state at $t\to -\infty$ \cite{saito_2006}. Sun {\it et al.}$~$compared the dynamics with and without the rotating-wave approximation (RWA) assuming the initial state of
the oscillator to be a superposition of coherent states, laying bare the inaccuracy of RWA \cite{sun_2012,Sun_2016}. Setting the harmonic oscillator in an initial
finite-temperature thermal equilibrium state, Ashhab considered the final probabilities of the LZSM transition \cite{ashhab_2014}. Huang and Zhao found two-stage LZ transitions
induced by the combined effect of tunneling strength $\Delta$ and the off-diagonal qubit-oscillator coupling \cite{huang2018}. Malla {\it et al.}$~$aimed to find to an analytical
solution in the presence of a slow and fast oscillator ($\omega < \Delta$ and $\omega > \Delta$, respectively, where $\omega$ is the frequency of the harmonic oscillator) assuming
there are many quanta excited initially \cite{MaRa18}. The same initial condition
has also been studied by Werther {\it et al.} \cite{Werther_2019_JCP}. Though many efforts have been devoted to understanding the LZSM model, several fundamental issues remain unsettled, such as the influence of an initially excited environment on the final {LZ} transition probability.
Moreover, the off-diagonal qubit-oscillator coupling has not been adequately treated if
{the oscillator is initialized in an non-vacuum state} \cite{sun_2012, ashhab_2014, MaRa18}. In this work, we will shed some light on the issues mentioned by investigating {the effects of}
initial superposition of coherent states of the oscillator {on} the LZSM dynamics of a two-level system coupled to a single harmonic oscillator using the time-dependent variational principle {combined with the multiple $\rm D_2$-{\it Ansatz}}.

The remainder of the paper is structured as follows. In Sec.~\ref{methodology}, we present the Hamiltonian and our trial wave function, the multi-$\mathrm{D}_{2}$ {\it
Ansatz}. {The observables of interest are described in Sec.~\ref{Observables}.}
In Sec.~\ref{Result_A}, we study the LZ model with an initial vacuum photon state and driven by a linear external field to illustrate the main physical picture of the
transition. In Sec.~\ref{initial_state}, it is demonstrated how to use
the Schr\"odinger-cat states to initialize the photon state in the framework of the multi-$\mathrm{D}_{2}$ {\it Ansatz}.
We then examine the detailed dynamics of the LZ model in the presence of a linearly varying field in Sec.~\ref{YS_state}, and in the presence of a sinusoidal driving field in Sec.~\ref{interferometer}. Special attention has been paid to the photon dynamics in a
setup similar to a LZSM interferometer. Conclusions are drawn in Sec.~\ref{Conclusion}.

\section{METHODOLOGY}
\label{methodology}

\subsection{A qubit coupled to a single mode}

\begin{figure}[tbp]
\centering
\includegraphics[clip,scale=0.3]{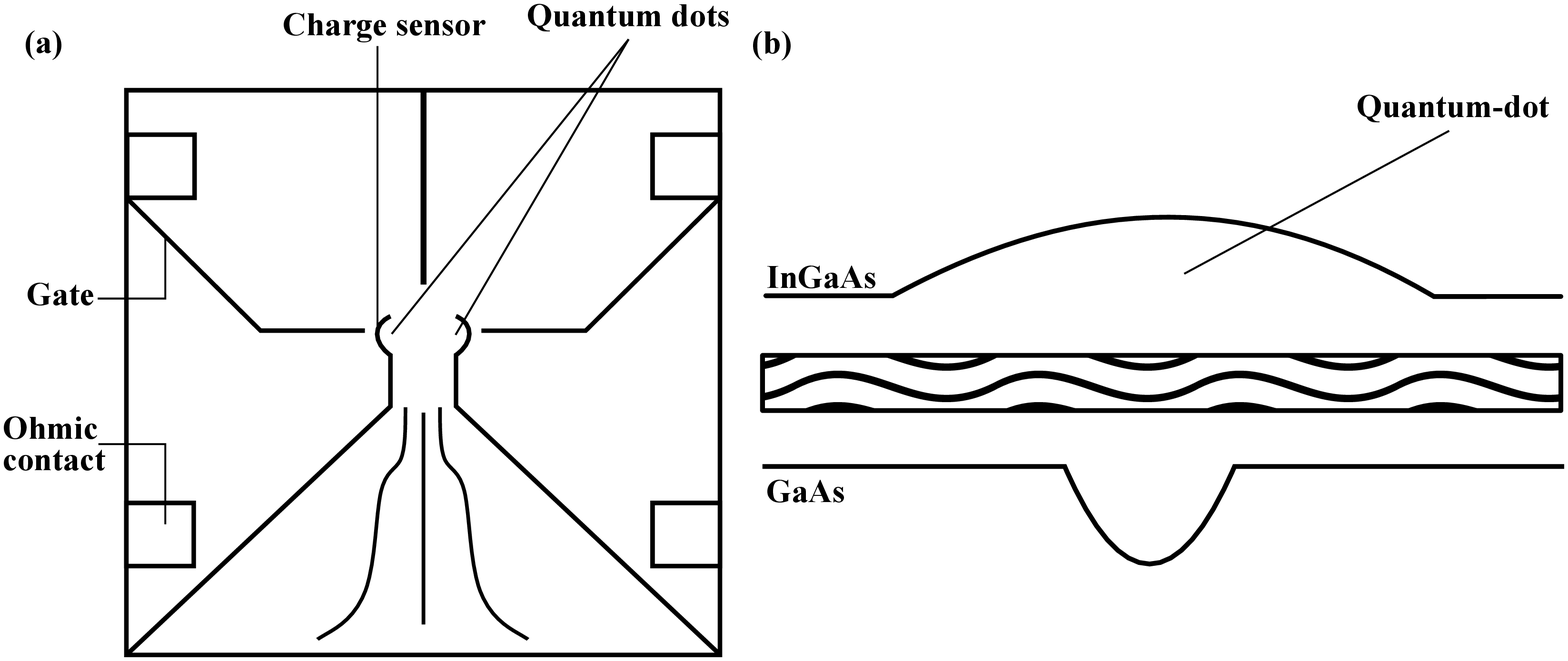}
\caption{Quantum dot and solid-state dopant qubits. (a) an electrostatically confined quantum dot; the structure shown is several $\mu \text{m}$ across. (b), a self-assembled quantum dot. }
\label{fig1}
\end{figure}

 The LZ transitions can be utilized in various quantum devices. One example that employs the LZ transition is adiabatic quantum computation.
Supposing a Hamiltonian of interest, $H_{\mathrm{f}}$, is difficult to implement in experiment. One can {then} construct a time-dependent Hamiltonian
 \begin{equation}
   H(t) = c_{\mathrm{i}}(t) H_{\mathrm{i}} + c_{\mathrm{f}}(t) H_{\mathrm{f}},
 \end{equation}
in which $H_{\mathrm{i}}$ is the initial Hamiltonian that can easily be implemented. $c_{\mathrm{i}}(t)$ and $c_{\mathrm{f}}(t)$ are functions depending
on time $t$ satisfying
$c_{\mathrm{i}}(0)=c_{\mathrm{f}}(T)=1$ and $c_{\mathrm{i}}(T)=c_{\mathrm{f}}(0)=0$, respectively. The target Hamiltonian $H_{\mathrm{f}}$ can be obtained if one increases the time
$t$ adiabatically \cite{yang_20}. The LZ transition can also help manipulate qubits and fabricate reliable readout devices \cite{Mason_2015}. Typical quantum devices can be subdivided into three classes,
the flux qubits, the charge qubits and the phase qubits \cite{Shevchenko_2010}.
Fig.~\ref{fig1}(a) supplies the time-dependent flux treading the quantum dots. Fig.~\ref{fig1}(b) displays the schematic
diagram of a superconducting quantum-dot coupled to a coplanar transmission line resonator \cite{Ladd_2010_Nat}.
LZ transitions occur frequently in the first two types of the devices \cite{Shevchenko_2010}. Meanwhile,
electromagnetic waves are used to readout the state of the qubit. In recent years the implementation of qubits interacting with an alternating electromagnetic field in
strong and ultrastrong regimes became possible \cite{huan_2020}. Stronger coupling leads to fast and reliable control of the qubits.

The total Hamiltonian of a driven two-level system interacting with a bosonic bath is given by
\begin{equation}
\hat{H}=\hat{H}_{\mathrm{S}}+\hat{H}_{\mathrm{B}}+\hat{H}_{\mathrm{SB}},
\label{Hamiltonian}
\end{equation}
where the system Hamiltonian is the standard LZSM Hamiltonian for an isolated two-level system, {i.e}, $\hat{H}_{\mathrm{S}}=\hat{H}_{\mathrm{LZSM}}$, with
\begin{equation}
\hat{H}_{\mathrm{LZSM}}=\frac{\varepsilon(t)\hbar}{2}\sigma_{z}+\frac{\Delta\hbar}{2}\sigma_{x},
\label{Hamiltonian_LZ}
\end{equation}
where $\sigma_{x}$ and $\sigma_{z}$ are the Pauli matrices. Denoted as diabatic states $\left|\uparrow\right\rangle$ and $\left|\downarrow\right\rangle$, the eigenstates of the qubit Hamiltonian
{$[\varepsilon(t)\hbar/2]\sigma_{z}$} have a time-dependent bias $\varepsilon(t)$. The tunneling strength $\Delta$ represents
intrinsic interactions between the diabatic states, and induces transitions between these states.

With an alternating external driving field, the time-dependent bias is written as
\begin{equation}
  \label{sin_driven}
  \varepsilon(t) = \varepsilon_{0} + A \sin (\Omega t + \varphi_{0}),
\end{equation}
in which $\varepsilon_{0}$ is the intrinsic bias of the qubit, $A$ is the driving amplitude, $\Omega$ is the frequency and $\varphi_{0}$ is the initial phase of the driving.

To study the LZ transition in the presence of an environment, we model a bosonic bath of one quantum harmonic oscillator by the Hamiltonian $\hat{H}_{\mathrm{B}}$
\begin{equation}
\hat{H}_{\mathrm{B}} = \hbar\omega\hat{b}^{\dagger}\hat{b}, \label{Hamiltonian_Bath}
\end{equation}
and couple the oscillator to the qubit via the Hamiltonian $\hat{H}_{\mathrm{SB}}$ \cite{WSKHK06}
\begin{equation}
\hat{H}_{\mathrm{SB}} = \frac{\gamma\hbar} {2}\left(\cos\theta_{\mathrm{c}}\sigma_{z}+\sin\theta_{\mathrm{c}}\sigma_{x}\right)(\hat{b}^{\dagger}+\hat{b}), \label{Hamiltonian_coupling}
\end{equation}
where $\gamma$ and $\theta_{\mathrm{c}}$ are the qubit-oscillator coupling and the interaction angle, respectively. $\omega$ indicates the frequency of the bath mode with
creation (annihilation) operator $\hat{b}^{\dagger}(\hat{b})$. The effect of the bosonic bath is to change the energies of the qubit via the diagonal coupling ($\sigma_z$) and to
induce transitions between the levels of the qubit via the off-diagonal coupling ($\sigma_{x}$). The interaction angle $\theta_{\mathrm{c}}=\pi / 2$ is assumed throughout this work.

In the presence of sinusoidal driving, the Hamiltonian reads
  \begin{equation}
    \label{one_mode_sin_H}
    \hat{H} = \frac{\varepsilon(t)\hbar}{2} \sigma_{z}+\frac{\Delta\hbar}{2} \sigma_{x}+ \hbar\omega \hat{b}^{\dagger} \hat{b}+\frac{\gamma\hbar}{2} \sigma_{x}\left(\hat{b}^{\dagger}+\hat{b}\right).
  \end{equation}
If the frequency $\Omega$ in Eq.~(\ref{sin_driven}) is rather small, one can approximate the sinusoidal driving
by a linear one. Then this system can be simply modeled by the Hamiltonian,
  \begin{equation}
    \label{one_mode_H}
    \hat{H}=\frac{v t}{2} \sigma_{z}+\frac{\Delta\hbar}{2} \sigma_{x}+ \hbar\omega \hat{b}^{\dagger} \hat{b}+\frac{\gamma\hbar}{2} \sigma_{x}\left(\hat{b}^{\dagger}+\hat{b}\right).
  \end{equation}
in which $v = A\Omega\hbar$ is the level-crossing speed. In this work, the tunneling $\Delta$ is set to
zero and $\hbar=1$. In Hamiltonian (\ref{one_mode_H}), there are then three parameters: the level-crossing speed $v$, the frequency of the photon $\omega$ and the coupling
strength $\gamma$. One can arbitrarily select one of the parameters as the characteristic parameter. In this article, we choose $\omega$ as the characteristic parameter and set it to unity throughout the paper. The unit
of the energy, time and speed are $\omega$, $\omega^{-1}$ and $\omega^{2}$, respectively.
With the parameters in Hamiltonian (\ref{one_mode_H}), one can construct a dimensionless parameter $\gamma^{2}/v$. The meaning of the dimensionless parameter will be revealed in
Sec.~\ref{Result_A}.

\subsection{The Multi-$\textrm{D}_2$ state}
\label{Multiple Davydov trial state}
Multiple Davydov trial states with multiplicity $M$ are $M$ copies of the corresponding single Davydov {\it Ansatz} \cite{zh_12,zh_97,Werther_2019_JCP}. The initial impetus of
their development was to investigate the full quantum dynamics of the polaron model \cite{zhou2015polaron, Zhou_16, huang_2017, huang_2017_off} and the spin-boson model
\cite{Wang_16, huang_SF_2017} in the framework of the time-dependent variational principle, putting the system and bath dynamics on an equal footing. In the two-level system, one of the
multiple Davydov trial states, the multi-$\textrm{D}_{2}$ {\it Ansatz} with multiplicity $M$, can be constructed as
\begin{eqnarray}
\label{D2_state}
&&\left|D_{2}^{M} (t)\right\rangle =\sum_{i=1}^{M}\left\{ A_{i}(t)\left|\uparrow\right\rangle \exp{\left[f_{i}(t)\hat{b}^{\dagger}-\mathrm{H.c.}\right]}\left|0\right\rangle \right\}\nonumber\\
&&+\sum_{i=1}^{M}\left\{B_{i}(t)\left|\downarrow\right\rangle \exp{\left[f_{i}(t)\hat{b}^{\dagger}-\mathrm{H.c.}\right]}\left|0\right\rangle \right\},
\end{eqnarray}
where $\mathrm{H.c.}$ denotes the Hermitian conjugate, and $\left|0\right\rangle$ is the vacuum state of the bosonic bath. $A_{i}$ and $B_{i}$ are time-dependent variational
parameter for the amplitudes in states $\left|\uparrow\right\rangle$ and $\left|\downarrow\right\rangle$, respectively. $f_{i}(t)$ are the bosonic displacements, where $i$
labels the $i$-th superposition state. If $M=1$, the multi-$\textrm{D}_{2}$ {\it Ansatz} is restored to the usual Davydov $\rm D_2$ trial state.

Equations of motion of the variational
parameters $u_{i}=$$A_{i}, B_{i}$ and $f_{i}$ are then derived by adopting the time-dependent variational principle,
\begin{equation}\label{eq:eom1}
\frac{d}{dt}\left(\frac{\partial L}{\partial\dot{u_{i}^{\ast}}}\right)-\frac{\partial L}{\partial u_{i}^{\ast}} =  0.
\end{equation}
For the multi-$\rm D_2$ {\it Ansatz}, the Lagrangian $L$ reads
\begin{eqnarray}
L& = & \langle {\rm D}^M_2(t)|\frac{i}{2}\frac{\overleftrightarrow{\partial}}{\partial t}- \hat{H}|{\rm D}^M_2(t)\rangle \nonumber \\
& = & \frac{i}{2}\left[ \langle {\rm D}^M_2(t)|\frac{\overrightarrow{\partial}}{\partial t}|{\rm D}^M_2(t)\rangle - \langle {\rm
D}^M_2(t)|\frac{\overleftarrow{\partial}}{\partial t}|{\rm D}^M_2(t)\rangle \right] \nonumber \\
&-& \langle {\rm D}^M_2(t)|\hat{H}|{\rm D}^M_2(t)\rangle.
\label{Lagrangian_2}
\end{eqnarray}
Details of the Lagrangian and equations of motion are given in the Supporting Information.

The exact diagonalization (ED) method has been adopted in Ref.~\cite{sun_2012} to solve similar problems. However, the ED method is only suitable for small systems each
  electronic/spin degree of freedom weakly coupled to a few boson modes. For electronic/spin systems strongly coupled to multiple boson modes, it is difficult for ED calculations
  to converge. Computationally, ED is expensive to implement due to the huge Hilbert space it has to cover, which is a result of large truncation values needed to describe the Fock
  states of the boson modes.  Moreover, the ED method is not suitable for dynamics at finite temperatures. In current study, we are interested in LZ transitions with the photon
  state initiated from some superpositions of coherent states. It is much more straightforward to use the Davydov {\it Ansatz} to estimate the wave functions because of the
  coherent-state construct of the boson portion of the Davydov {\it Ansatz}. Meanwhile, the approach applied in this work is quite powerful to deal with the Schr\"odinger-cat states
  with increasing photon displacements, which would be difficult for the ED method to handle due to the increasing Hilbert space dimensions. In the future, we plan to study
  temperature effects on the LZ transitions in systems with the qubit coupled to multiple photon modes. It would be impossible for the ED method to solve these problems. Therefore,
  we apply the time-dependent variational method with the Davydov {\it Ansatz} from the very beginning of the project. The ED method has been only used to provide some benchmarks for some simple cases.

\section{Results and discussion}

\subsection{Observables}\label{Observables}
Initially the qubit is in its up state $\left |\uparrow \right\rangle$. To investigate the LZ transition quantitatively, the probability that the qubit in the down state $\left |\downarrow \right\rangle$ is usually employed and is denoted as $P_{\mathrm{LZ}}(t)$. With
the density matrix $\rho (t)= \left | D_{2}^{M}(t) \right\rangle  \left \langle D_{2}^{M}(t) \right | $, $P_{\mathrm{LZ}}(t)$, can be expressed as
\begin{equation}
\label{observable_PLZ}
\mathrm{P}_{LZ}(t)=\mathrm{Tr} [\rho (t) \hat{P}_{\downarrow} ] =  \frac{1}{2}\left[1-\left\langle\hat{\sigma}_{z} \right\rangle (t) \right].
\end{equation}
Here $\hat{P}_{\downarrow} \equiv \left|\downarrow \right\rangle \left\langle\downarrow \right|$
is the projection operator of the down state. With the help of the multi-$\mathrm{D}_2$ {\it Ansatz}, the expectation of the Pauli operator $\sigma_{z}$ can be expressed as
$\left\langle\hat{\sigma}_{z}\right\rangle (t) = \left\langle\mathrm{D}_{2}^{M} (t) \left|\hat{\sigma}_{z}\right| \mathrm{D}_{2}^{M} (t) \right\rangle$.

In order to explore the detailed transfer pathways, we also record the population dynamics in coupled qubit-photon states
$\left | n,\, \uparrow \right\rangle \equiv \left | \uparrow \right \rangle \bigotimes \left | n \right \rangle$ and $\left | n,\, \downarrow \right\rangle
  \equiv \left | \downarrow \right \rangle \bigotimes \left | n \right \rangle $ where $\left | n \right \rangle$ is the photon Fock state. Using the projection operators
$\hat{P}_{n,\uparrow} \equiv \left | \uparrow \right \rangle  \left \langle \uparrow \right | \bigotimes \left | n \right \rangle \left\langle n \right | $ and $\hat{P}_{n,\downarrow} \equiv \left | \downarrow \right \rangle  \left \langle \downarrow \right | \bigotimes \left | n \right \rangle \left\langle n \right | $, we define the population as
  \begin{equation}
    \label{up_fockstates}
    P_{n,\uparrow } = \mathrm{Tr} ( \rho \hat{P}_{n,\uparrow}),
  \end{equation}
 for state $\left | n,\, \uparrow \right\rangle $, and
  \begin{equation}
    \label{observable_fockstates}
    P_{n,\downarrow } = \mathrm{Tr} ( \rho \hat{P}_{n,\downarrow}),
  \end{equation}
for state $\left | n,\, \downarrow \right\rangle$.
These quantities are evaluated with the multi-$\mathrm{D}_2$ {\it Ansatz},
\begin{equation}
  \label{Fock_population}
    P_{n,\uparrow (\downarrow)} = {\left| \left\langle n,\, \uparrow (\downarrow)  | D_{2}^{M} \right\rangle  \right|}^{2}.
\end{equation}

\subsection{Physical understanding of Landau-Zener processes}
\label{Result_A}
\begin{figure}[tbp]
  \includegraphics[scale= 0.3]{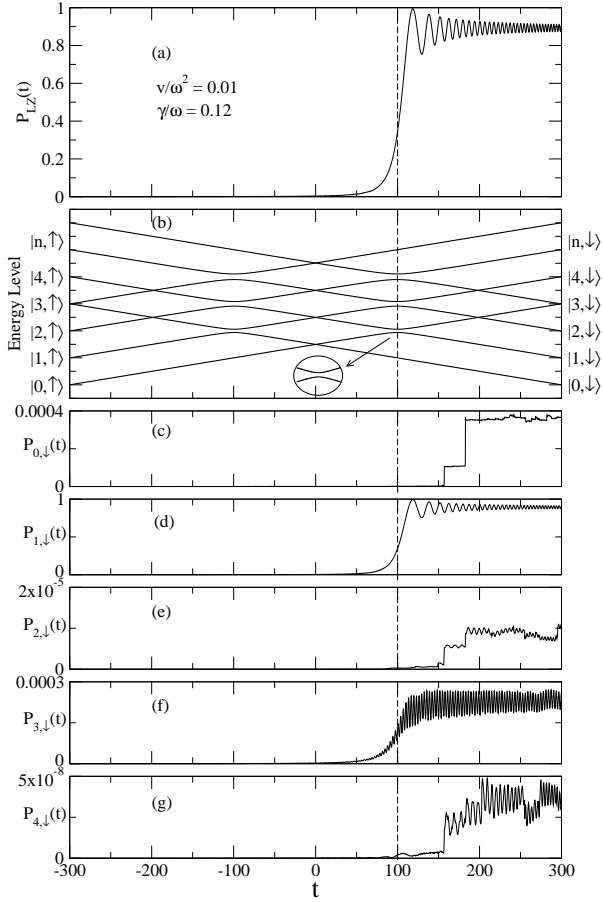}\\
  \caption{
      (a) Time evolution of the transition probability calculated by the multi-$\mathrm{D}_{2}$ {\it Ansatz}. (b) The eigenstate diagram of Hamiltonian (\ref{one_mode_H}). The
      inset is a zoomed view of an avoided crossing between the states $\left|0,\, \uparrow \right\rangle$ and  $\left|1,\, \downarrow \right\rangle$ at $t=100$. (c)-(g) Time evolution of the population
      $P_{n,\downarrow}$ in the state $|n,\, \downarrow \rangle $ with $n=0$, 1, ..., 5. Level-crossing speed $v/\omega^2=0.01$, other parameters are $\Delta=0$  and
      $\gamma/\omega = 0.12$. }
\label{fig2}
\end{figure}

In this subsection, we will discuss the change of the energy levels with the time in the LZ system with a linearly driving field. The parameters that we choose are the
level-crossing speed $v/\omega^2=0.01$, and the coupling strength {$\gamma/\omega = 0.12$}. Diagonalizing the Hamiltonian Eq.~(\ref{one_mode_H}) directly, the lowest several energy
levels varying with time are obtained and are illustrated in {Fig.~\ref{fig2}(b)}. As shown in {Fig.~\ref{fig2}(b)}, the energy of the states $\left |n,\, \uparrow \right \rangle$ rises
with increasing time, and the energy of the states $\left |m,\, \downarrow \right \rangle $ decreases with time. Especially, at {$t = 100$}, the avoided
crossing between $\left|0,\, \uparrow\right\rangle$ and  $\left|1,\, \downarrow \right\rangle$ is zoomed in the subfigure. In the vicinity of {$t = 100$}, the system evolves into the
superposition state $\alpha (v)\,\left|0,\, \uparrow \right\rangle +\beta (v)\,\left|1,\, \downarrow \right\rangle$ with velocity-dependent probability amplitudes. The dynamics of the model is
dominated by the energy levels shown in the diagram, if the coupling strength is weak and the external field varies adiabatically.

To illustrate how $P_{\mathrm{LZ}}(t)$ depends on the energy levels, simulations are performed for Hamiltonian (\ref{one_mode_H}). {With the same parameters as in Fig.~\ref{fig2}(b), $P_{\mathrm{LZ}}(t)$ is plotted in Fig.~\ref{fig2}(a)}. The vacuum initial condition of the bath is employed. As shown in Fig.~\ref{fig2}(a), the
probability $P_{\mathrm{LZ}}$ is always close to 0 from $t=-300$ to $t=0$. But when the time $t$ approaches $100$, the probability suddenly surges to
$P_{\mathrm{LZ}}(t) = 0.9$. With increasing time, the value of the probability oscillates, and approaches a stable value gradually. To ensure convergence, several multiplicities
($M =6,\ 8,\ 10$) are used, and curves with different multiplicities are found to coincide perfectly, as shown in Figure \ref{figS1} in the Supporting Information.

The dynamics of $P_{\mathrm{LZ}}(t)$ can be understood from the energy levels in Fig.~\ref{fig2}(b). At $t=100$, an
avoided crossing occurs, as revealed in Fig.~\ref{fig2}(b). Meanwhile, in {Fig.~\ref{fig2}(a)}, one can find that $P_{\mathrm{LZ}}(t)$ suddenly jumps up at the same time $t=100$,
which is indicated by a vertical dashed line. The entire system is initialized at the $\left|0,\,  \uparrow \right\rangle $ state.
In the adiabatic limit, the system will end up in the $\left|1,\, \downarrow \right\rangle $ state, based on the
adiabatic theorem.
The qubit flips in the vicinity of the avoided crossing. In Hamiltonian (\ref{one_mode_H}), the avoided crossing results from the off-diagonal
coupling term $\gamma / 2 \sigma_{x}(\hat{b}^{\dagger}+\hat{b})$, which can be considered as dynamical tunneling.
The parameters we used here satisfy that $\gamma / \omega = 0.12$ and {$\gamma^{2}/v = 1.44$}.
If $\gamma \ll \omega$ and $v \lesssim \gamma^{2}$, the model is approximately in the weak coupling regime, and the
external field varies adiabatically with time. In this regime, the dynamics of $P_{\mathrm{LZ}}(t)$ near the avoided crossing can be predicted from the energy levels shown in
Fig.~\ref{fig2}(b). Based on the adiabatic theorem, for a vanishingly small speed $v$, the system initially at $\left|\uparrow\right\rangle$ will be finally at $\left|\downarrow\right\rangle$
  when $t$ approaches infinity. On the other hand, the coupling strength $\gamma$ widens the gap of the avoided crossing. Thus, $v$ increases the likelihood the LZ transition while
  $\gamma$ suppresses it, and the dimensionless parameter $\gamma^{2}/v$ combines the two aspects.

To investigate the LZ transition from the perspective of the bath \cite{Fujihashi_2017}, the populations of the five lowest down states ${P}_{n,\downarrow }(t)$ are plotted in Fig.~\ref{fig2}(c)-(g).
Except for ${P}_{1,\downarrow }(t)$, all other ${P}_{n,\downarrow }(t)$ have vanishingly small amplitudes.
${P}_{1,\downarrow }(t)$ suddenly rises at $t=100$. From the aforementioned discussion, the
qubit flip is from the avoided crossing between $\left|0,\, \uparrow \right\rangle $ and $\left|1,\, \downarrow \right\rangle $. The same information can be extracted from
Fig.~\ref{fig2}(c)-(g) as well. It is also obvious that for the given vacuum initial photon state in this subsection, the only relevant final state is the $n=1$ photon Fock state. In
Fig.~\ref{fig2}(d), after the jump near $t=100$, $P_{1,\downarrow}(t)$ fluctuates around 0.88. The parameters selected to obtain the curves in {Fig.~\ref{fig2}(a) and
(c)-(g)} satisfy $vt \gg \gamma$ and $\omega \gg \gamma/2$. The contribution of the off-diagonal coupling term is proportional to $\gamma$, while the contribution of Hamiltonian
(\ref{Hamiltonian_LZ}) is proportional to the time-dependent bias $\varepsilon=vt$. Thus, away from the LZ transition point, the coupling between the qubit and the
bath is a perturbation to Hamiltonian (\ref{one_mode_H}). It follows that the system will remain in $\left|1,\, \downarrow \right\rangle$ at long times.

The LZ transition emerges out of multi-level interactions in the time-dependent qubit-photon system.
As shown in {Fig.~\ref{fig2}}, probing the origins of the LZ transitions and attributing them to relevant avoided crossings is an interesting task in the
dynamics analysis. Approaches that are based on the reduced density matrix of the qubit while tracing out the photon degree of freedom are unable to reveal the bath dynamics explicitly. As a wave function-based method, the multi-$\mathrm{D}_{2}$ {\it Ansatz} can readily reveal the detailed bath dynamics, a feature that is unavailable to approaches based on density
matrices.

\subsection{Landau-Zener transition with an initial superposition of coherent states}\label{initial_state}

Superpositions of coherent states, such as the Schr\"{o}dinger-cat states, have attracted extensive interest as nonclassical states with exceptional
properties. Such states can be prepared in various systems, essential in many fundamental tests of quantum theory and in myriad quantum-information-processing tasks
\cite{Zurek_2003,Ourjoumtsev_2006,Ourjoumtsev_2007}, including quantum computation \cite{lund2008fault}, precision measurements \cite{joo2011erratum,afek2010high}, and quantum
teleportation \cite{van2001entangled}.
Here, our aim is to study the effect of the initial Schr\"odinger-cat states on the LZ transition.

A coherent state $|\alpha\rangle$ is usually expanded as an infinite sum of Fock state $|n\rangle$, and the expansion coefficients for each Fock state decrease with the increasing
photon number $n$ [cf.~Eq.~(\ref{number_state1})]. With an initial photon state composed of coherent states, such as the
Yurke-Stoler (YS) state \cite{yurke1986generating}, the time-dependent variational principle with the Davydov {\it Ans\"atze} is a very handy wave-function propagation tool for the simulation task. Moreover, in the evolution process, the external
driving field may change the energies of the two-level system dramatically. Through the interaction between the qubit and the photon, the photon field may be excited to the state
with a large average photon number. Methods based on the Fock state will be less efficient than those based on coherent states, and less accurate as well due to the necessary
truncation of the Hilbert space.

\begin{figure*}[tbp]
\centering
\includegraphics[scale=0.45]{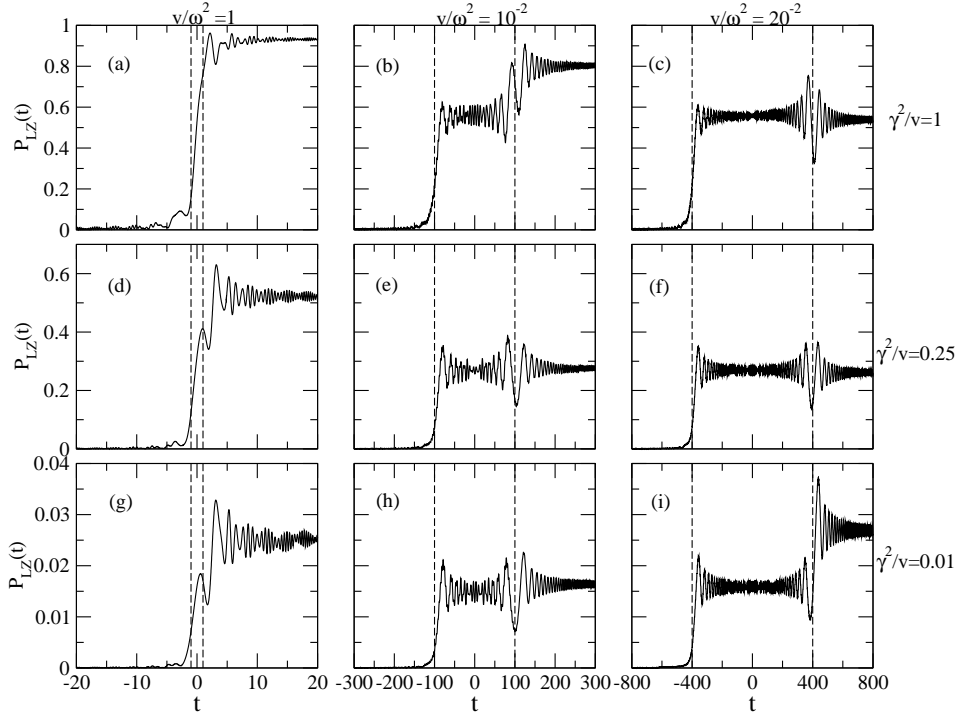}
\caption{Time (in units of $\omega^{-1}$) evolution of the transition probability from the multi-$\mathrm{D}_2$ {\it Ansatz} with dimensionless parameter
  $\gamma^{2}/ v$ and speed $v$ (in units of $\omega^{2}$). The vertical black dashed lines indicate the times of the avoided-level crossings.
  The photon displacement $|\alpha| = 1$. In (a), (d), and (g), $v/\omega^2 = 1$; in
    (b), (e), and (h), $v/\omega^2=10^{-2}$; and in (c), (f), (i), $v/\omega^2=20^{-2}$.  }
\label{fig3}
\end{figure*}

Next, we will focus on the propagation from a Schr\"odinger-cat state, $\left| \Psi(t= 0, \theta)\right\rangle_{\rm ph}$, defined as
\begin{equation}
\label{coherent_states}
\left| \Psi(t= 0, \theta)\right\rangle_{\rm ph} = \frac{1}{N_{\theta}}(\left| \alpha\right\rangle + e^{i\theta}\left| -  \alpha\right\rangle ),
\end{equation}
where $\left| \alpha\right\rangle$ is a coherent state  with displacement $\alpha$, $\theta$ is the phase and ${{N}_{\theta }}^{2}=2(1 + {e}^{-2|\alpha|^2 } \cos \theta)$ is the normalization
constant. For simplicity, we assume $\alpha$ to be real. If the phase $\theta$ is 0, $\pi$ and $\pi/2$, the state is named the even coherent state, the odd coherent state and the YS state \cite{yurke1986generating}, respectively.

Coherent states are the quasi-classical states of a quantum harmonic oscillator. But superpositions of coherent states are much more complex. The Mandel $Q$ parameter
  \begin{equation}
  \label{mandel_q}
    Q = \frac{\langle (\hat{n}-\langle \hat{n} \rangle )^{2} \rangle - \langle \hat{n} \rangle} {\langle \hat{n} \rangle } = \frac{\langle \hat{n}^{2} \rangle -\langle \hat{n} \rangle ^{2}
    - \langle n \rangle } {\langle \hat{n} \rangle},
\end{equation}
is usually used to measure the departure of the occupation number distribution from Poisson statistics. Here $\hat{n} = b^{\dagger}b$ is the boson number operator, and
$\langle \hat{n} \rangle$ and $\langle \hat{n^{2}} \rangle$ are the average of $\hat{n}$ and $\hat{n}^{2}$, repectively \cite{mandel_1979}. For the Schr\"odinger-cat states, the Mandel $Q$ parameters depend on the phase $\theta$. For the even state, the YS state and the odd state, $Q > 0$, $Q = 0$ and $Q < 0$, respectively. Thus, the even state is classical, the YS state is quasiclassical and the odd state is non-classical, respectively.

Many-body dynamics resulting from an initial superposition of photon coherent states can be more complex than that from an initial
vacuum state. This can be understood qualitatively from Fig.~\ref{fig2}(b). A coherent state can be written as a sum of Fock states \cite{shankar2012principles}
  \begin{equation}
    \label{number_state1}
    |\alpha \rangle ={{e}^{-\frac{|\alpha {{|}^{2}}}{2}}}\sum\limits_{n=0}^{\infty }{\frac{{{\alpha }^{n}}}{\sqrt{n!}}}|n\rangle.
  \end{equation}
  With an initial coherent state, for example, the initial population will be distributed to multiple $\left|n,\, \uparrow\right\rangle$ states with $n=$ 0, 1, 2, $\dots$, and the entire wave packet will be spread out over many $\left|n,\, \uparrow\right\rangle$ and $\left|n,\, \downarrow\right\rangle$ states. Thus, as the population in state $\left|0,\, \uparrow \right\rangle$ migrates to other states in the vicinity of the avoided crossing, the same would happen to higher
  excited states $\left|n,\, \uparrow \right\rangle$. Meanwhile, the expansion coefficients ${e}^{-|\alpha |^{2} / {2}}  (\alpha^{n} / \sqrt{n!}) $ lends a weightage to
    $\left|n,\, \uparrow \right\rangle$.
  For an initial superposition state given by {Eq.~(\ref{coherent_states})}, the ensuing dynamics also depends on the phase $\theta$. In the next subsection, we will
  investigate the transition probability $P_{\mathrm{LZ}}(t)$ from an initial photon state given by {Eq.~(\ref{coherent_states})} with various values of $\alpha$ and
  $\theta$.

The even coherent state can be written as,
  \begin{equation}
    |\alpha {{\rangle }_{\rm even}}=\frac{(|\alpha \rangle +|-\alpha \rangle )}{N_0}=\frac{{e}^{-\frac{|\alpha {{|}^{2}}}{2}}}{N_0}  \sum\limits_{n=0}^{\infty }{\frac{{{\alpha }^{2n}}}{\sqrt{2n!}}}|2n\rangle, \nonumber
  \end{equation}
  and the odd coherent state,
  \begin{equation}
    |\alpha {{\rangle }_{\rm odd}}=  \frac{(|\alpha \rangle -|-\alpha \rangle )} {{{N}_{\pi }}}
    = \frac{e^{-\frac{|\alpha {{|}^{2}}}{2}}}{N_\pi}  \sum\limits_{n=0}^{\infty }{\frac{{{\alpha }^{2n+1}}}{\sqrt{(2n+1)!}}}|2n+1\rangle, \nonumber
  \end{equation}
  where $N_0=2(1+e^{-2|\alpha|^2})$ and $N_\pi=2(1-e^{-2|\alpha|^2})$ are the normalization factors for the even and the odd coherent state, respectively.
  $ {{\left| \alpha  \right\rangle }_{\rm even}} = \left| \Psi(0, 0)\right\rangle_{\rm ph} $ is
named ``even coherent state" because it is an even parity state, and the photon number distribution is nonzero only for even photon numbers with an average photon number of $2|\alpha {{|}^{2}}(1-{{e}^{-2|\alpha {{|}^{2}}}})/N_{0}^{2}$. Similarly, $ {{\left| \alpha  \right\rangle }_{\rm odd}} = \left| \Psi(0, \pi)\right\rangle_{\rm ph} $ is
named ``odd coherent state" because it is an odd parity state, and the photon number distribution is nonzero
only for an odd number of photons with an average photon number of $2|\alpha {{|}^{2}}( 1+{{e}^{-2|\alpha {{|}^{2}}}} )/N_{\pi }^{2}$. Lastly,
if $\theta =\pi /2$, the initial photon state is the YS coherent state, i.e.,
${{\left| \alpha  \right\rangle }_{\mathrm{YS}}}=\left| \Psi(0, \pi/2)\right\rangle_{\rm ph} = (\left| \alpha\right\rangle +i\left| -\alpha  \right\rangle )/{N}_{\pi /2}$, with
an average photon number of $|\alpha {{|}^{2}}$ {and $N_{\pi/2}^{2}=2$}.

Compared with this work, an extra term $\omega \sigma_{z}/2$, where $\omega$ is the frequency of the photon bath, is included in the Hamiltonian of Ref.~\cite{sun_2012}, for which Sun {\it et al.}~obtained the LZ transition probability at long times,
\begin{equation}
  \label{state_probability}
  P_{\mathrm{LZ}}(\infty) = 1 - \frac{2P_{\uparrow,0}} {N_\theta^2e^{|\alpha|^2}} (e^{|\alpha|^2P_{\uparrow,0}} + e^{-|\alpha|^2P_{\uparrow,0}} \cos\theta).
\end{equation}
Here, $P_{\uparrow,0}=\exp(-\pi \gamma^2/2v)$ is the probability staying at $\left|\uparrow\right\rangle$ if the initial state is $\left|0,\, \uparrow\right\rangle$.
The RWA has been employed to derive this result. Although the probability is dependent on the detailed Hamiltonian, Eq.~(\ref{state_probability}) yields a common physical picture
of the system dynamics in the weak coupling regime. From Eq.~(\ref{state_probability}), one can calculate the possibilities for the even, odd and YS states. It follows that for
even coherent states, the final RWA transition probability is
\begin{equation}
\label{even_state_probability}
P_{{\rm LZ, even}}(\infty) = 1 - P_{\uparrow,0}\dfrac{\cosh(|\alpha|^{2}P_{\uparrow,0})}{\cosh|\alpha|^{2}},
\end{equation}
and for odd coherent states,
\begin{equation}
\label{odd_state_probability}
P_{\mathrm{LZ, odd}}(\infty)=1-P_{\uparrow, 0} \frac{\sinh \left(|\alpha|^{2} P_{\uparrow, 0}\right)}{\sinh |\alpha|^{2}}.
\end{equation}
As $|\alpha {{|}^{2}}$ vanishes, one has $P_{\mathrm{LZ, odd}}(\infty )\to 1-P_{\uparrow ,0}^{2}$, because the odd coherent state $|\alpha {{\rangle }_{\rm odd}}$ approaches the Fock state $|1 \rangle $. Similarly, for the YS coherent state, the final RWA transition probability is
\begin{equation}
\label{YS_state_probability}
P_{LZ}(\infty) = 1 - P_{\uparrow,0}\exp[-|\alpha|^{2}(1-P_{\uparrow,0})].
\end{equation}
Eq.~(\ref{YS_state_probability}) reveals the dependence of $P_{\mathrm{LZ}}(\infty )$ on the ratio ${\gamma^{2}}/v$ and the average photon number $|\alpha {{|}^{2}}$. Obviously, enhancing $|\alpha {{|}^{2}}$ and the ratio ${\gamma^{2}}/v$ will increase the final LZ probability $P_{\mathrm{LZ}}(\infty )$.

To simulate the LZ dynamics with an initial photon state of Eq.~(\ref{coherent_states}), the multi-$\textrm{D}_{2}$ trial state of Eq.~(\ref{D2_state}) is initialized as follows. For the qubit amplitudes, we set $A_1=1$, $A_2=e^{i \theta}$, and $B_1=B_2=0$. To ensure numerical stability, the rest of qubit amplitudes, $A_{2m-1}$, $A_{2m}$, $B_{2m-1}$ and $B_{2m}$ ($m=2$, 3, 4, $\cdots$, $M/2$), are set to small random numbers between $- 10^{-4}$ and $+10^{-4}$.
We set photon displacements $f_1=\alpha$ and $f_2= -\alpha$. $f_{2m-1}=f_1$ and $f_{2m}=f_2$ ($m=2$, 3, 4, $\cdots$, $M/2$) at $ t = -\infty$.
It is obvious that the multiplicity of our multi-$\textrm{D}_{2}$ trial state should be an even number.
In the interest of numerical stability, additional random numbers no greater than $\pm 10^{-2}$, are added to $f_{2m-1}$ and $f_{2m}$ ($m=2$, 3, 4, $\cdots$, $M/2$).

\subsection{Linear driving}\label{YS_state}

In order to gain insight into LZ dynamics with an initial YS state at intermediate times, we numerically calculated LZ probabilities for several values of {the speed $v$ ($v=c \omega^{2}$, $c=1^{-2}$, $10^{-2}$ and $20^{-2}$) and the dimensionless parameter $\gamma^{2}/v$ ($ \gamma^{2}/v = 1^{2}$, $0.5^2$ and $0.1^{2}$). Here, based on the
  considerations below, we choose $\gamma^{2}/v$ rather than the coupling strength $\gamma$ as the control parameter. The probability term $P_{\uparrow,0}=\exp(-\pi \gamma^2/2v)$,
  which is determined by the ratio $\gamma^{2}/v$, influences many related probabilities in the LZ transitions, such as Eq.~(\ref{state_probability}) \cite{sun_2012}. More
  examples can be found in Ref.~\cite{saito_2007}. In the literature, the LZ model is in the fast driving regime if $v/\Delta^{2} \gg 1$, and in the slow driving regime if
  $v/\Delta^{2} \ll 1$ \cite{Arceci_2017_PRB,Chen_2020_PRB}. {As the off-diagonal coupling strength $\gamma$ determines dynamical tunneling of the LZ transition, one
  can use $v/\gamma^{2}$ to ascertain whether the driving is fast or slow.}
For various values of $v$ and $v/\gamma^{2}$, simulations are performed to reveal the dynamics of $P_{\mathrm{LZ}}(t)$ in different regimes. Meanwhile, panels in the same row of Fig.~\ref{fig3} have the same $\gamma^{2}/v$.} Displayed in
Figs.~\ref{fig3} (b), (c), (e), (f), (h) and (i) are the transition probabilities in the weak coupling regime as $\gamma \ll \omega$ is satisfied.
There are two distinct stages of LZ transitions in Figs.~\ref{fig3} (b), (c), (e), (f), (h) and (i), which can be extracted from {the eigenstate diagram similar to} Fig.~\ref{fig2} (b).

As shown in Fig.~\ref{fig2} (b) , there are two series of avoided crossings at $t=\pm \omega / v$, leading to two successive transitions.
As displayed in Figs.~\ref{fig3} (b), (e), (h) and (c), (f), (i),
the height of the first transition is
insensitive to the speed $v$ but is dependent on $\gamma^{2}/v$. This behavior can be understood with the help of Eq.~(\ref{YS_state_probability}). At long times, the RWA transition probability $P_{\rm LZ}(\infty)$ is determined by $P_{\uparrow,0}=\exp(-\pi\gamma^2/2v)$, and therefore by $\gamma^{2}/v$.
This asymptotic behavior obtained with the
RWA is valid only in the weak coupling regime.
At a sufficiently long time from the first transition, the evolution of $P_{\mathrm{LZ}}(t)$ obtained with the RWA coincides
with that of Hamiltonian (\ref{one_mode_H}) before the onset of the second transition.
Thus, the height of the first jump in the transition probability $P_{\mathrm{LZ}}(t)$
only depends on $\gamma^{2}/v$ in the weak coupling regime.

In Figs.~\ref{fig3} (b), (e) and (h), a second jump in the probability $P_{\mathrm{LZ}}(t)$ occurs at about {$t=100$}, while in Figs.~\ref{fig3} (c), (f) and (i), it happens at {$t=400$}. It is found that these times correspond to the avoided crossings at $t=\omega/v$. As shown clearly in Figs.~\ref{fig3} (b) and (i), the height
of the plateau after the second jump in $P_{\mathrm{LZ}}(t)$ deviates from that of the first jump. It follows that the second transition cannot be described by RWA. The avoided crossing at $t=\pm \omega / v$ can be divided into two classes. The
first class is formed between $\left|n,\, \uparrow \right\rangle$ and $\left|n+1,\, \downarrow \right\rangle$, and the second, between $\left|n,\, \uparrow \right\rangle$ and $\left|n-1,\, \downarrow \right \rangle$. As the second class of the avoided crossings can not be described in any way by the RWA, the RWA fails for the LZ transition at long times, as supported by our numerical results in Fig.~\ref{fig3}. Figs.~\ref{fig3} (a) and (d) illustrate the LZ transition probability in the strong coupling regime, and Fig.~\ref{fig3}(g) displays a marginal case sandwiched  between the strong
and the weak coupling regime. Only one LZ transition is found in Figs.~\ref{fig3} (a), (d), and (g). The avoided crossings at $t=\pm \omega / v$ still exist,
but the speed of $v/\omega^2=1$ renders too small the temporal separation of two LZ transitions between the two avoided crossings, therefore merging two transitions into one.
In the simulations shown in Fig.~\ref{fig3}, the photon displacement $|\alpha| = 1$. According to Fig.~\ref{fig2} (b), the second class of transitions from $\left|n,\,
    \uparrow\right\rangle$ to $\left|n-1,\, \downarrow\right\rangle$ is feasible for $n\geqslant3$ at the first set of avoiding crossings, $t = -\omega/v$. Thus, if a sufficiently
  large $|\alpha|$ is used, the RWA will break down, a conclusion that is supported by additional numerical simulations performed.

\begin{figure}
\centering
\includegraphics[scale=0.3]{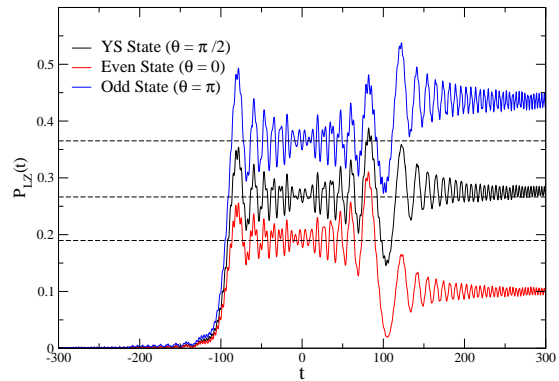}
\caption{{Time evolution of the transition probability calculated by the multi-$\textrm{D}_{2}$ {\it Ansatz} for different initial states. The speed is $v/\omega^{2} = 0.01$, and the coupling strength is $\gamma / \omega =0.05$.}}
\label{fig4}
\end{figure}

\begin{figure}
  \centering
  \includegraphics[scale=0.3]{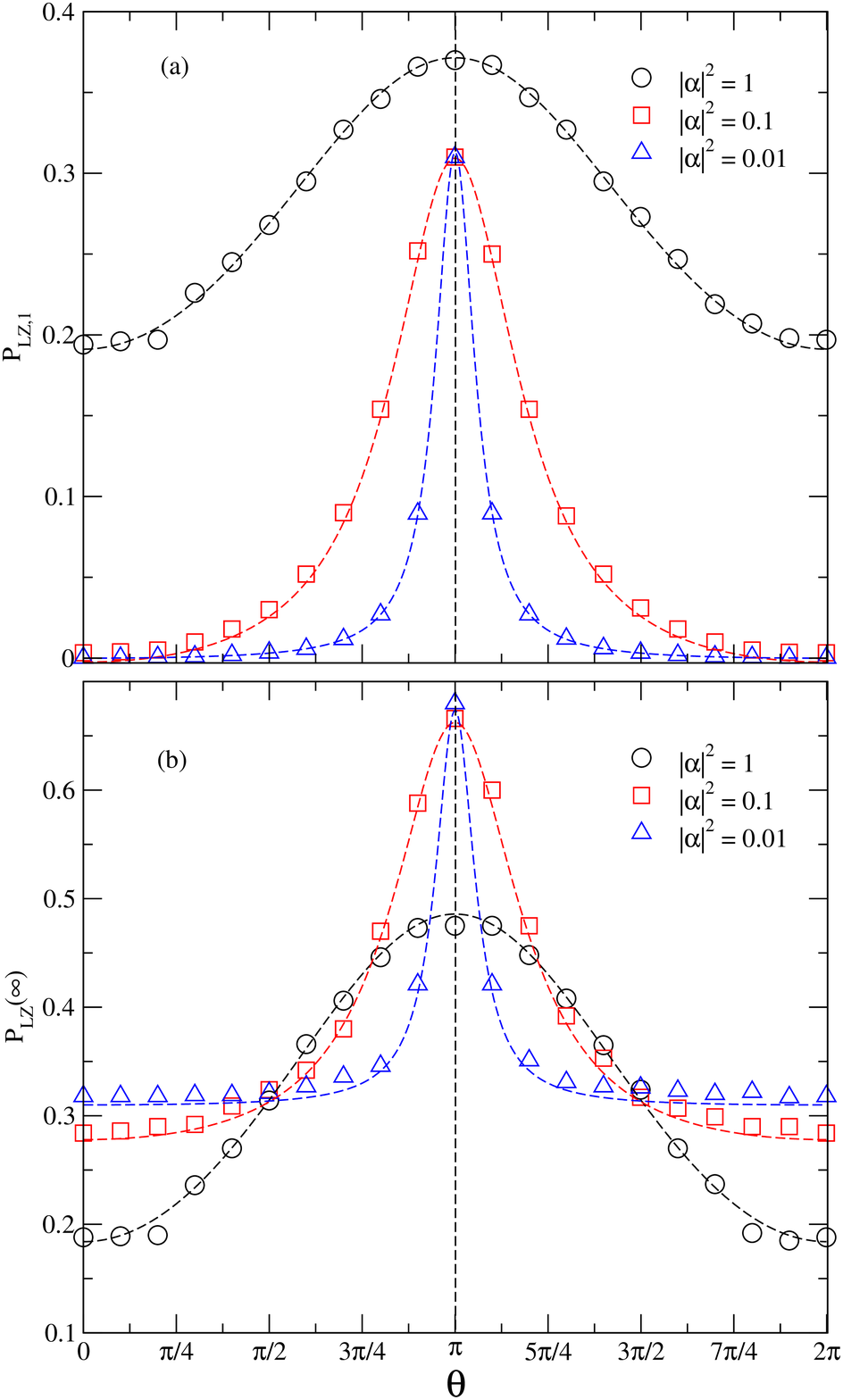}\\
  \caption{The average height of transition probabilities of the two consecutive transitions as functions of phase $\theta$ in
      Eq.~(\ref{coherent_states}) with $|\alpha|^{2}=1$ (black circles), 0.1 (red squares) and 0.01 (blue triangles). Fitting curves with Eq.~(\ref{fit_equation}) are plotted with
        black dashed lines for $|\alpha|^{2}=1$, red dashed lines for $|\alpha|^{2}=0.1$, and blue dashed lines for $|\alpha|^{2}=0.01$.
        (a): The average height of the first plateau, $P_{\mathrm{LZ},1}$. (b): The height of the plateau after the second transition $P_{\mathrm{LZ}}(\infty)$.
       {The speed is $v/\omega^{2}=0.01$, and the coupling strength is $\gamma/\omega =0.05$.} }
\label{fig5}
\end{figure}

To further investigate how the phase $\theta$ influences the transition probability $P_{\mathrm{LZ}}(t)$, simulations are performed for $\theta=0$, $\pi$ and $\pi/2$, i.e., the even coherent state, the odd coherent state, and the YS state, respectively. As shown in Fig.~\ref{fig4}, the three curves for different values of the phase $\theta$ exhibit almost the same oscillatory behavior, with
the first and the second transitions occurring at about the same times.
The height of the first plateaus rises with the increasing $\theta$. This trend agrees with the prediction of the RWA transition probabilities given in Eqs.~(\ref{even_state_probability}), (\ref{YS_state_probability}) and (\ref{odd_state_probability}). As discussed earlier, the second transition in  $P_{\mathrm{LZ}}(t)$ cannot be described by the RWA. Our simulation results here shed light on how the phase $\theta$ affects
the probability $P_{\mathrm{LZ}}(\infty)$.
To compare the heights of the first and second plateaus, dashed lines are drawn to display the average positions of the first plateaus. As shown in Fig.~\ref{fig4}, the second plateau is found to be slightly lower than the first for the even coherent state,  while for the YS state and the odd coherent state, the opposite is true.
It is clear from our simulations that to increase the $P_{\mathrm{LZ}}(\infty)$, one has to increase the phase $\theta$ all the way up to $\pi$.

Fig.~\ref{fig5} displays the transition probabilities of the aforementioned two consecutive transitions as functions of the phase $\theta$ for $|\alpha|^2=1$, 0.1,
  0.01, {$v/\omega^2=0.01$, and $\gamma/\omega=0.05$.}
 The average height of the first plateau, $P_{\mathrm{LZ, 1}}$, as a function of $\theta$ is plotted in Fig.~\ref{fig5}(a). The symbols (circles, squares, and triangles) are extracted from
 the simulation with multi-$\textrm{D}_{2}$ trial states. $P_{\mathrm{LZ,1} } $ increases (drops) as $\theta$ goes from $0$ ($\pi$) to $\pi$ ($2\pi$). Asides from minute numerical
 fluctuations, the symbols (circles, squares, and triangles) in Fig.~\ref{fig5}(a) have mirror symmetry about $\theta = \pi$, which can be fitted by
\begin{equation}
\label{fit_equation}
P = F_{0} - \frac{1+e^{2|\alpha|^{2}F_{1}} \cos \theta} {1+e^{2|\alpha|^{2}} \cos \theta} e^{|\alpha|^{2}F_{1}}
\end{equation}
a relation that resembles Eq.~(\ref{state_probability}).
The fitting parameters are $F_{0}=2.68$ and $F_{1}=0.88$ for $|\alpha|^{2}=1$, $F_{0}=1.1$ and $F_{1}=0.72$ for $|\alpha|^{2}=0.1$, and $F_{0}=1.01$ and $F_{1}=0.69$ for
$|\alpha|^{2}=0.01$. The fitting curves are plotted as the colored dashed lines in Fig~\ref{fig5}(a). Although Eq.~(\ref{state_probability}) was derived from a slightly different
Hamiltonian by Sun {\it et al.} in \cite{sun_2012}, the fitting in Fig.~\ref{fig5}(a) seems to be satisfactory.

In Fig.~\ref{fig5}(b), the height of the plateau as the system re-equilibrates after the second transition, $P_{\mathrm{LZ}}(\infty)$, is shown as (circles, squares or triangles) a function of the phase
  $\theta$, which also displays mirror symmetry about $\theta = \pi$. One can also fit $P_{\mathrm{LZ}}(\infty)$ with Eq.~(\ref{fit_equation}) with different values of the parameters $F_{0}$ and $F_{1}$.
The fitting parameters are $F_{0}=2.68$ and $F_{1}=0.88$ for $|\alpha|^{2}=1$, $F_{0}=1.1$ and $F_{1}=0.72$ for $|\alpha|^{2}=0.1$, and $F_{0}=1.01$ and
  $F_{1}=0.69$ for $|\alpha|^{2}=0.01$, respectively. A comparison of Fig.~\ref{fig5}(a) and Fig.~\ref{fig5}(b) reveals how the displacement $|\alpha|^{2}$ influences the plateau heights.
The height of the plateau after the first transition increases with the increasing $|\alpha|^{2}$ for any phase $\theta$.
But the situation is much more complex for the second plateaus. As shown in Fig.~\ref{fig5}(b),
$P_{\mathrm{LZ}}(\infty)$ decreases with increasing $|\alpha|^{2}$ near $\theta=0$, $\pi$ and $2\pi$. But between about $\theta=\pi/2$ ($5\pi/4$) and $3\pi/4$ ($3\pi/2$),
$P_{\mathrm{LZ}}(\infty)$ increases with the increasing $|\alpha|^2$.

\subsection{Sinusoidal driving }\label{interferometer}

{Beyond linear driving, periodical external fields are also frequently adopted to drive the qubits, making such a simple model an ideal platform to investigate various fundamental
  physical problems. For instance, Schr\"{o}dinger-cat states have been generated via photon-assisted LZSM interferometry by repeatedly sweeping the energy splitting of the qubit \cite{Lidal_2020}. Recently it has also been found that the states of qubits and photons in a complex  Rabi-dimer system can be engineered by driving the qubits with sinusoidal fields \cite{Huang_2019_JCP, Zheng_2021_JCP}.}
Detailed dynamics related to those multiple LZ transitions has not been well understood, as it is a nontrivial task to analyze the coupled qubit-photon dynamics at LZ transitions. In this section, we present an in-depth analysis of the qubit-photon dynamics at multiple LZ transitions by depicting the population dynamics in individual coupled qubit-photon states. Transition pathways are revealed by combining the energy diagram of the coupled qubit-photon states and the total energy.

\begin{figure}
  \centering
  \includegraphics[scale=0.3]{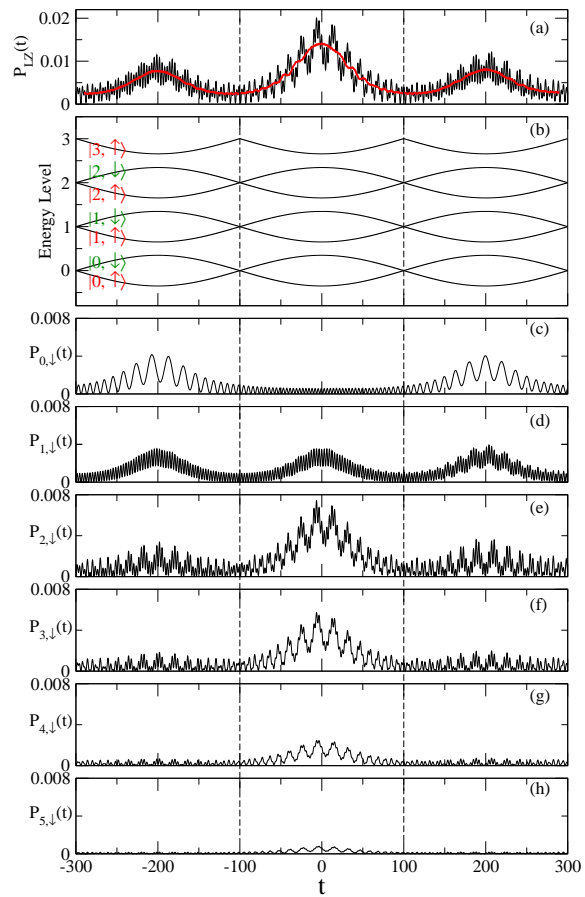}\\
  \caption{(a) Time evolution of the transition probability $P_{\rm LZ}(t)$ calculated by the multi-${\rm D}_{2}$ {\it Ansatz}. {The red line is a moving average of
      $P_{\mathrm{LZ}}(t)$ over $\Delta t = 200$.} (b) The eigenstate diagram of Hamiltonian (\ref{one_mode_sin_H}). (c)-(h) Time evolution of the population
      $P_{n\downarrow}(t)$ in the state $\left|n,\, \downarrow\right\rangle$  with $n=0$, 1, $\dots$, 5.
    The initial condition of Eq.~(\ref{coherent_states}) is adopted with phase $\theta=\pi/2$ and displacement $|\alpha|^{2}=1$. Parameters in Eq.~(\ref{sin_driven}) for sinusoidal driving are $\varepsilon_{0}=0$, $A=0.7$, $\Omega/\omega=\pi/200$, $\varphi_{0}=\pi/2$. The coupling strength is $\gamma/\omega=0.05$. }
\label{fig6}
\end{figure}

As described in Eq.~(\ref{sin_driven}), a sinusoidal external driving field is imposed to the qubit with a driving frequency $\Omega/\omega = \pi/200$, a coupling strength $\gamma/\omega = 0.05$, a initial phase $\varphi_{0} = \pi/2$, and $\varepsilon_{0} = 0$.
The superposition state $|\alpha\rangle_{\mathrm{YS}}$ with the displacement $|\alpha|^{2}=1$ and the phase $\theta=\pi/2$ is adopted as the initial photon state.
Three driving amplitudes, i.e., $A = 0.7$, $1.1$, and $1.3$, are used in the simulations.

Driven by a sinusoidal field with an amplitude of $A=0.7$, the qubit migrates from the initial up state $\left|\uparrow\right\rangle$ to the down state
$\left|\downarrow\right\rangle$ with a probability $P_{\mathrm{LZ}}$. As illustrated by the black line in Fig.~\ref{fig6} (a), the transition probability $P_{\mathrm{LZ}}$
oscillates as a function of time with three characteristic frequencies, which can be seen more clearly in the moving average of $P_{\mathrm{LZ}}(t)$ over a time interval $\Delta t
= 200$ (the red line in Fig.~\ref{fig6} (a)). Low frequency oscillations have a period of 200, which is half of the driving period, i.e., $T_{\rm d} = 2\pi/\Omega = 400$. Those
oscillations in $P_{\mathrm{LZ}}(t)$ arise from tunnelings between the $\left|n,\, \uparrow \right\rangle$ and
$\left|n-1,\, \downarrow \right\rangle$ ($\left|n+1,\, \downarrow \right\rangle$) states. These tunnelings can be well understood with the help of the energy diagram of the hybrid
qubit-photon system as shown in Fig.~\ref{fig6} (b), which is obtained by diagonalizing Hamiltonian (\ref{one_mode_H}). For instance, the
$\left|n,\, \uparrow \right\rangle$ $\rightarrow$ $\left|n-1,\, \downarrow \right\rangle$ (for $n\geqslant$1) tunnelings occur around $t=-200$ and $200$, while the $\left|n,\,
  \uparrow \right\rangle \rightarrow \left|n+1,\, \downarrow \right\rangle$ tunnelings, in the vicinity of $t=0$. As the driving amplitude $A=0.7$ is smaller than the photon
frequency $\omega$, there is no level crossing between qubit-photon states with different photon numbers. Therefore, only a minor portion of population tunnels to the qubit down
state at the times when the $\left|n,\, \uparrow \right\rangle$ and $\left|n-1,\, \downarrow\right\rangle$ ($\left|n+1,\, \downarrow \right\rangle$) states have small energy
gaps. The time-dependent energy gaps give rise to the oscillations in $P_{\mathrm{LZ}}(t)$ with time-dependent periods. In order to clarify these oscillations, we present the
population dynamics on all qubit-photon states in
Figs.~\ref{fig6} (c)-(h). The population on $\left|0,\, \downarrow \right\rangle$ is an excellent example to elaborate the oscillations with time-dependent periods as shown in
Fig.~\ref{fig6} (c). The population on $\left|0,\, \downarrow \right\rangle$ is tunneled from $\left|1,\, \uparrow \right\rangle$. Comparing the energy levels of these two states in
Fig.~\ref{fig6} (b), it is found that the time-dependent energy gap agrees perfectly with the oscillation frequency in Fig.~\ref{fig6} (c). For example, the energy
gap arrives at its minima at $t=-200$ and $200$, leading to the maximum oscillation periods in $P_{0,\downarrow}$. At $t=0$, the two states have a maximum energy gap, giving rise
to the fast oscillations in the population on $\left|0,\, \downarrow \right\rangle$. The third characteristic oscillation frequency in $P_{\mathrm{LZ}}(t)$ dependents on the
frequency of the photon mode and is $2\omega$. Independent of time, this frequency is  responsible for the fastest oscillations in $P_{\mathrm{LZ}}(t)$.

Analyzing the energy levels in Fig.~\ref{fig6} (b) and the population dynamics on all qubit-photon states presented in Fig.~\ref{fig6} (c)-(h), we can reveal the detailed pathways
for the tunnelings taking place along time. At the initial time, more than 90\% of the population is distributed on $\left|0,\, \uparrow \right\rangle$,
$\left|1,\, \uparrow \right\rangle$, and $\left|2,\, \uparrow \right\rangle$. The population on $\left|0,\, \uparrow \right\rangle$ can only tunnel to
$\left|1,\, \downarrow \right\rangle$ at $t=0$. All the other $\left|n,\, \uparrow \right\rangle$ ($n \geqslant 1$) states can
tunnel their population to $\left|n+1,\, \downarrow \right\rangle$ at $t=0$, and to $\left|n-1,\, \downarrow \right\rangle$ at $t=-200$ and $200$. Therefore, as recipients, all the
$\left|n,\, \downarrow \right\rangle$ states except $\left|0,\ \downarrow \right\rangle$ exhibit three local maxima in their populations at $t=-200$, $0$, and $200$. As discussed
above, $\left|0,\ \downarrow \right\rangle$ only receives the tunneling from $\left|1,\ \uparrow \right\rangle$ at $t=-200$, and $200$, producing two local maxima in $P_{0,\downarrow}$ at $t=-200$, and $200$.

\begin{figure}[tbh]
  \centering
    \includegraphics[scale=0.3]{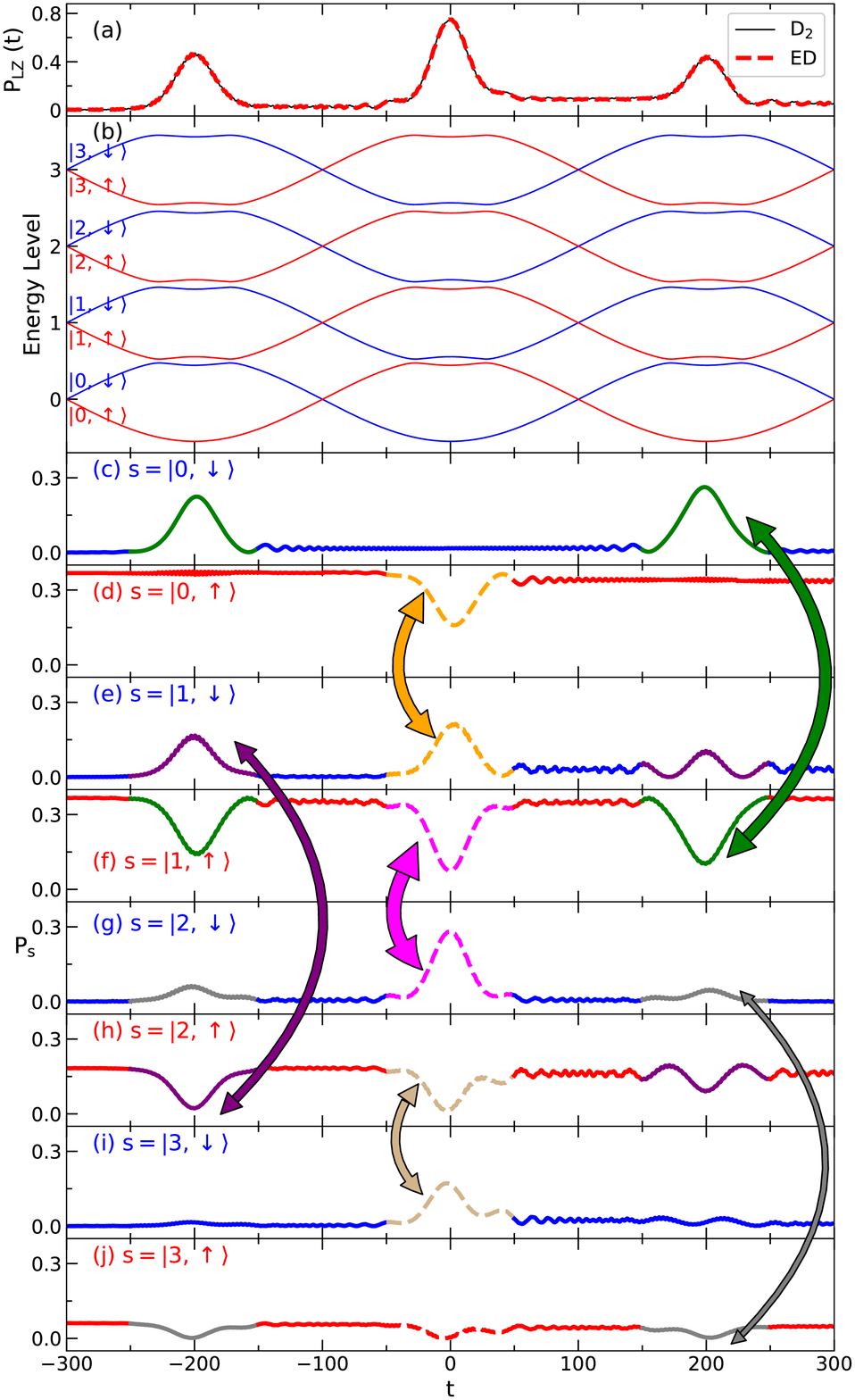}
  \caption{(a) Time evolution of the transition probability $P_{\rm LZ}(t)$ calculated by the multi-${\rm D}_{2}$ {\it Ansatz}. {The solid line is $P_{\mathrm{LZ}}(t)$ from the
    multi-$\mathrm{D}_{2}$ {\it Ansatz}, and the dashed line is $P_{\mathrm{LZ}}(t)$ from exact diagonalization. }
    (b) The eigenstate diagram of Hamiltonian (\ref{one_mode_sin_H}). (c)-(j) Time evolution of the
      photon bath. The transitions between states are labeled by curved arrows. The initial condition of Eq.~(\ref{coherent_states}) is adopted with phase $\theta=\pi/2$ and displacement $|\alpha|^{2}=1$. Parameters in Eq.~(\ref{sin_driven}) for sinusoidal driving are $\varepsilon_{0}=0$, $A=1.1$, $\Omega/\omega=\pi/200$, $\varphi_{0}=\pi/2$. The
      coupling strength is $\gamma/\omega=0.05$.}
\label{fig7}
\end{figure}

With a larger driving amplitude of $A=1.1$, avoided crossings appear between qubit-photon states with different photon numbers, as illustrated by the energy diagram in Fig.~\ref{fig7}
(b). At these avoided crossings, a series of photon-induced LZ transitions take place, giving rise to the prominent
$\left|\uparrow \right\rangle$ $\rightarrow$ $\left|\downarrow \right\rangle$ transition probability shown in Fig.~\ref{fig7} (a). In the vicinity of the avoided crossings, the
energy gaps between the $\left|\uparrow \right\rangle$ and $\left|\downarrow \right\rangle$ states are quite small, leading to low-frequency oscillations in the population, such as
the two oscillations of $P_{0,\downarrow}$ at $t=-200$ and $200$ in Figs.~\ref{fig7} (c). It is clear that the period of these low-frequency oscillations is comparable
to (or even larger than) the interval between the two avoided crossings surround the oscillation. Therefore, only three main peaks emerge in the down-state population  (Fig.~\ref{fig7} (a)).

Similar to the aforementioned weak driving case, detailed mechanisms for LZ transitions can be unveiled by analyzing the population dynamics on individual coupled qubit-photon
states, as illustrated in Figs.~\ref{fig7} (c)-(j). The overall LZ transition probability shown in Fig.~\ref{fig7} (a) arises from a bunch of LZ transitions between different
qubit-photon states as labelled by the arrows in Figs.~\ref{fig7} (c)-(j). These transitions can be classified into two types. One type of these transitions occur around $t=-200$
and $200$, and follow the pathway $\left|n+1,\, \uparrow \right\rangle$ $\rightarrow$ $\left|n,\, \downarrow \right\rangle$ $\rightarrow$ $\left|n+1,\, \uparrow \right\rangle$. For
instance, the purple arrow labels the $\left|2,\, \uparrow \right\rangle \rightarrow \left|1,\, \downarrow \right\rangle$ transition at $t=-228$ and the back transition from
$\left|1,\, \downarrow \right\rangle$ to $\left|2,\, \uparrow \right\rangle$ at $t=-172$. Similar transitions are also labeled by the green and gray arrows around $t=200$. The other type of
transitions proceed from a pathway of $\left|n,\, \uparrow \right\rangle$ $\rightarrow$ $\left|n+1,\, \downarrow \right\rangle$ $\rightarrow$ $\left|n,\, \uparrow \right\rangle$, and
occur around $t=0$. Combining with the energy diagram in Fig.~\ref{fig7} (b), we find that both types of LZ transitions originate in the adiabatic evolution of the wave function
starting from some $\left|n,\, \uparrow \right\rangle$ states along the time-dependent eigenstates, i.e., the energy levels of these eigenstates are show in red in Fig.~\ref{fig7}
(b). For a given displacement $\alpha$, the initial population is distributed over several $\left|n,\, \uparrow \right\rangle$ states with different amplitudes, leading to different
contributions to the LZ transitions. The dominance of these contributions is presented by the amplitudes of the peaks (dips) in $P_{n,\downarrow} (t)$ ($P_{n,\uparrow} (t)$), and
are further highlighted by the width of the curved arrows in Figs.~\ref{fig7} (c)-(j). One can find that, the {dominant} transition around $t=-200$ and $200$ occurs between
$\left|1, \uparrow \right\rangle$ and $\left|0,\, \downarrow \right\rangle$. Nearby $t=0$, the transition between $\left|1,\, \uparrow \right\rangle$ and $\left|2,\, \downarrow
\right\rangle$ contributes the most to the overall transition probability variation.

After each of such $\left|\uparrow \right\rangle$ $\rightarrow$ $\left|\downarrow \right\rangle$ $\rightarrow$ $\left|\uparrow \right\rangle$ transitions, a small portion of population is
 accumulated on the down state, which can be seen from the following plateau of each peak. The energy gaps at avoided crossings are different for different pairs of adiabatic states due to off-diagonal
 qubit-photon coupling, i.e.,
 $\Delta E_{\left|3,\ \uparrow \right\rangle - \left|2,\ \downarrow \right\rangle} > \Delta E_{\left|2,\ \uparrow \right\rangle - \left|1,\ \downarrow \right\rangle}
> \Delta E_{\left|1,\ \uparrow  \right\rangle - \left|0,\ \downarrow \right\rangle}$. Hence, as can be seen from the central peaks in Figs.~\ref{fig7} {(c)-(j)}, shorter period oscillations are found in
populations of states with higher photon numbers at avoided crossings. As shown in Figure \ref{figS2} in the Supporting Information, there is a perfect linear relation between the inverse of the
energy gap $\Delta E_{|n+1,\ \uparrow \rangle - |n,\ \downarrow \rangle}$ at $t=-28$ and the oscillation period of $P_{n, \downarrow}$ ($n>0$) around $t=0$.
\begin{figure}[tbh]
  \centering
  \includegraphics[scale=0.3]{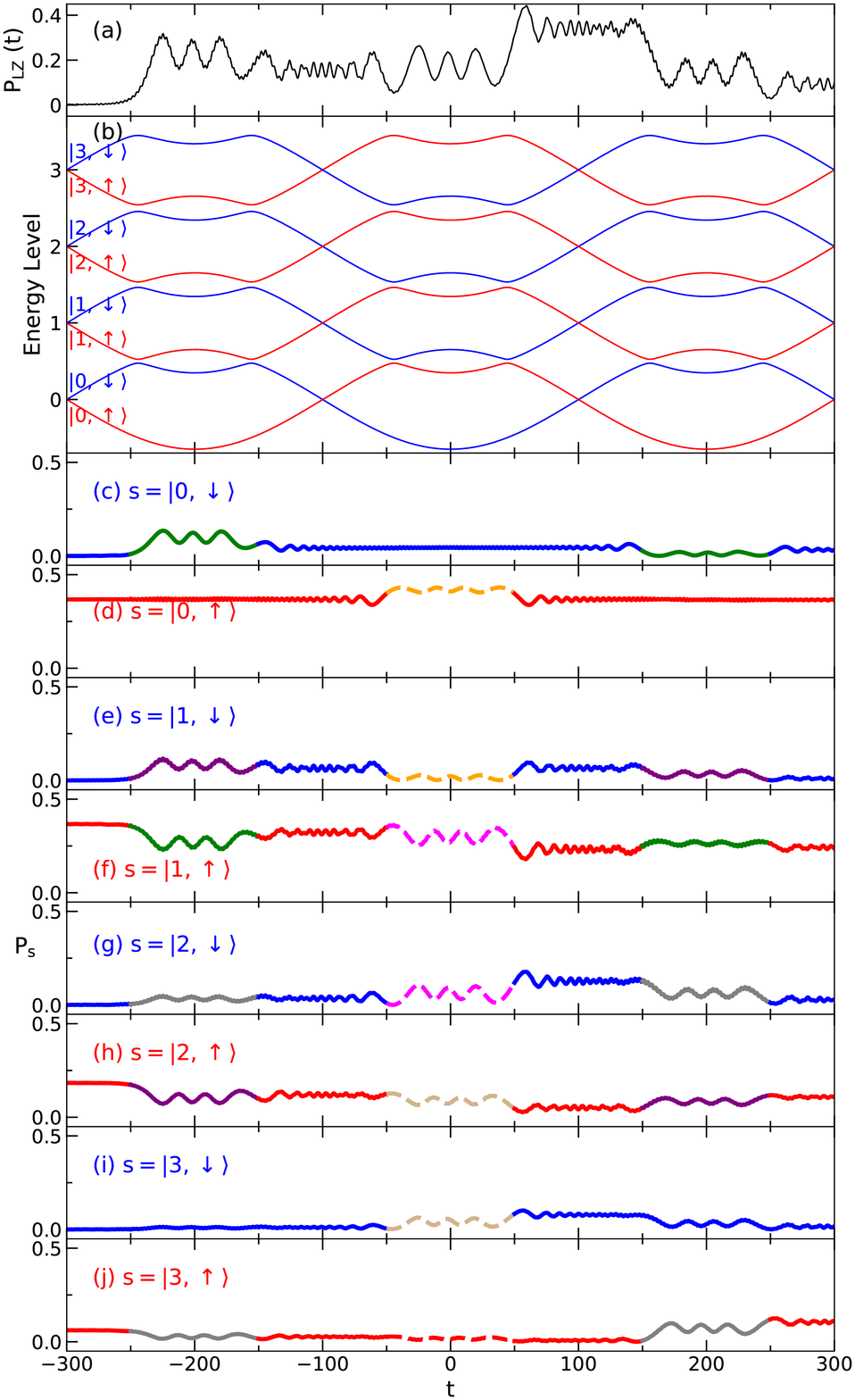}
  \caption{(a) Time evolution of the transition probability $P_{\rm LZ}(t)$ calculated by the multi-${\rm D}_{2}$
    {\it Ansatz}. (b) The eigenstate diagram of Hamiltonian (\ref{one_mode_sin_H}). (c)-(j) Time evolution of the photon bath. The initial condition of Eq.~(\ref{coherent_states}) is adopted with phase $\theta=\pi/2$ and
    displacement $|\alpha|^{2}=1$. The parameters in Eq.~(\ref{sin_driven}) for the sinusoidal driving is $\varepsilon_{0}=0$, $A=1.3$, $\Omega/\omega=\pi/200$, $\varphi_{0}=\pi/2$. The coupling
    strength is $\gamma/\omega=0.05$.}
 \label{fig8}
\end{figure}

As further confirmation of our efficient, wave-function based approach,
we compare the results from the multi-$\mathrm{D}_{2}$ {\it Ansatz} with those from exact diagonalization.
As shown in Fig.~\ref{fig7}(a), $P_{\mathrm{LZ}}(t)$ from the multi-$\mathrm{D}_{2}$ {\it Ansatz} (solid line) coincides
with that from the method of exact diagonalization (dashed
  line), demonstrating that time-dependent variation with the multi-$\mathrm{D}_{2}$ {\it Ansatz} is a reliable approach to the LZSM dynamics. Additional comparisons between the
  multi-$\mathrm{D}_{2}$ {\it Ansatz} and exact diagonalization for different initial photon states and external driving fields can be found in Figure S1 in the Supporting Information.

Increasing the driving amplitude further to $A=1.3$ produces quite different dynamics compared to aforementioned cases. The large driving amplitude leads to a large time-spacing
between two adjacent avoided crossings as shown in Fig.~\ref{fig8} (b). The population can reside on the down state for a longer time after each $\left|\uparrow \right\rangle$
$\rightarrow$ $\left|\downarrow \right\rangle$ transition. Between two adjacent avoided crossings, large energy gaps between two nearby adiabatic states lead to short oscillation periods in populations of the coupled qubit-photon states.
Therefore, plateaus with oscillations appear between the two avoided crossings as depicted in Fig.~\ref{fig8} (a),
in contrast to a single peak between two avoided crossings shown in Fig.~\ref{fig7} (a) with $A=1.1$. Nevertheless, the large energy gaps hinder the population transfer between the
two adiabatic states, but favor instead the diabatic population flow. In the vicinity of the avoided crossing, the driving is approximately linear and the driving speed is
  $v\sim A\Omega\hbar$. With an increasing amplitude $A$, the speed will increase and the external field sweeps through the avoided crossing more quickly. The system will have a high likelihood to evolve along the diabatic surfaces with a larger $A$.
As can be seen from the population dynamics on different coupled qubit-photon states shown in Figs.~\ref{fig8} (c)-(j), both adiabatic and diabatic evolutions of wave packets are involved in the dynamics of this scenario.

\begin{figure}
  \centering
  \includegraphics[scale=0.3]{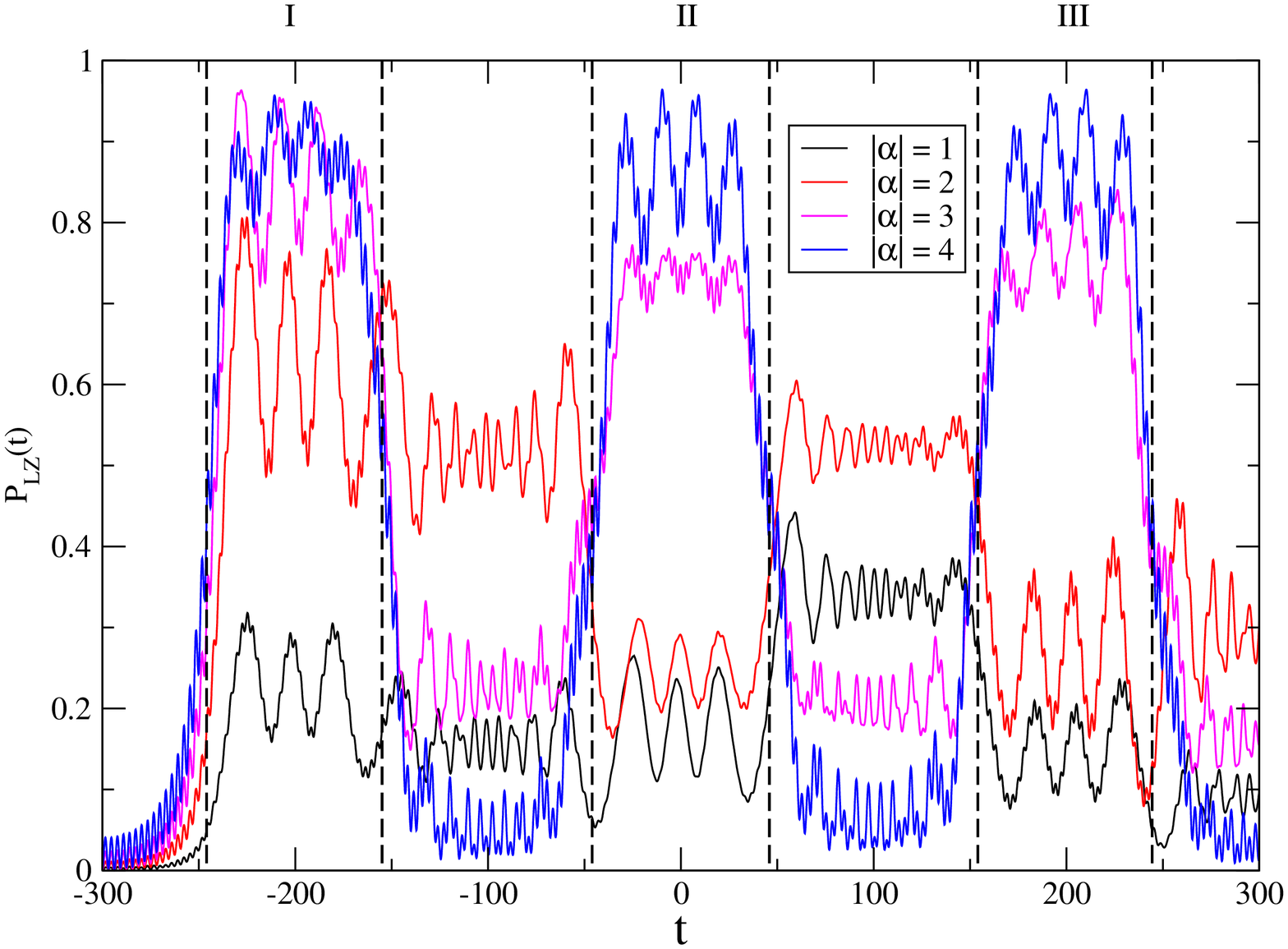}
  \caption{{Time evolution of the transition probability $P_{\rm LZ}(t)$ calculated by the multi-${\rm D}_{2}$
    {\it Ansatz}. The initial condition of Eq.~(\ref{coherent_states}) is adopted with phase $\theta=\pi/2$. Four amplitudes of the displacement $|\alpha|=1$(black), 2(red), 3(orange) and 4(blue) are employed. The parameters in
      Eq.~(\ref{sin_driven}) for the sinusoidal driving is $\varepsilon_{0}=0$, $A=1.3$, $\Omega/\omega=\pi/200$, $\varphi_{0}=\pi/2$. The coupling strength is $\gamma/\omega=0.05$.}}
\label{fig9}
\end{figure}

To probe how the average photon number $|\alpha|^{2}$ influences the dynamics, YS states $\left|\alpha \right\rangle_{\mathrm{YS}}$ with different amplitudes of displacement
$|\alpha|$ are used as the initial photon state. As shown in Fig.~\ref{fig9}, $|\alpha|$ equal to 1.0 (black), 2.0 (red), 3.0 (orange) and 4.0 (blue) are employed. Other parameters
are the same with as in Fig.~\ref{fig8}. The vertical dashed lines label the locations of the avoided crossings which are also shown in Fig.~\ref{fig8} (b). Two successive avoided
crossings form time intervals labeled with I, II and III. In Intervals I, II and III, one can find that $P_{\mathrm{LZ}}(t)$ increases with $|\alpha|$. A larger average photon
number yields larger $P_{\mathrm{LZ}}(t)$ in the intervals. This trend coincides with what is shown in Fig.~\ref{fig5} (a) for the linear driving. For each curve in
Fig.~\ref{fig9}, the difference of $P_{\mathrm{LZ}}(t)$ in and out the intervals increases with $|\alpha|$. A larger difference of $P_{\mathrm{LZ}}(t)$ implies that the system is
more likely staying at $\left|\uparrow\right\rangle$ after the second avoided crossing of an interval if the average photon number is larger. To understand this behavior, one can
consider Fig.~\ref{fig8} (b), where energy gaps at avoided crossings become larger with the increasing index of Fock state $n$. A larger average photon number means that an avoided
crossing between states $\left|n,\, \uparrow \right\rangle$ and $\left|n\pm1,\, \downarrow \right\rangle$ with a larger $n$ is involved, leading to a  larger energy gap that prevents the
transition form $\left|n,\, \uparrow \right\rangle$ to $\left|n\pm1 ,\, \downarrow \right\rangle$.

\section{Conclusion}\label{Conclusion}

In this work, extensive simulations have been performed for a driven qubit interacting
with a cavity photon mode which is initialized with a Schr\"odinger-cat state.
To arrive at numerically exact dynamics of the qubit and its photon bath in a LZ model with off-diagonal coupling, the
multi-$\mathrm{D}_{2}$ {\it Ansatz} has been utilized in combination of the time-dependent variational principle.
Linear and sinusoidal fields are employed to drive the qubit. For a linearly driven qubit, the problem is a photon-assisted LZ model, while for a sinusoidally driven qubit, one has a LZSM interferometer.

Previous investigations of the LZ transition often focus on the qubit dynamics while paying less attention to the evolution of the photon field.
As the LZ transition involves multiple energy levels in the presence of the photon mode, it is non-trivial to decipher which energy levels contribute to the transition if the initial photon state is not a Fock state. In this work,
with the help of the multi-$\mathrm{D}_{2}$ {\it Ansatz}, one can directly identify the individual contributions of the energy levels from the computed photon population in the up state (down state) $P_{n, \uparrow (\downarrow)}$. The effectiveness and validity of our method to extract individual contributions of the energy levels is first established in a photon-assisted LZ system with linear driving and an initial photon vacuum state.
Then our validated approach is applied to a qubit coupled with a photon mode and driven by a sinusoidal
external field, and the photon mode is initialized by the Schr\"odinger-cat state. It is found that within our method the LZSM transitions can be readily attributed to individual energy levels that are responsible.

Taking the Schr\"odinger-cat state as the initial photon state, we have explored
the photon-assisted LZ model both in the weak coupling regime ($\gamma < \omega$) and in the fast driving regime ($v/ \gamma^{2} \gg 1$), using the multi-$\mathrm{D}_{2}$ {\it Ansatz}.
Two consecutive transitions are uncovered in our numerical results. It has been claimed that for a LZ model with a constant bias, the first of the two transition can be described qualitatively by the RWA, but the second is beyond the reach of the RWA \cite{sun_2012}.
From our simulations, similar behavior is discovered in the absence of the
constant bias term. It is found that a constant bias only influences the detail of the dynamics but does not change the overall dynamical features of the LZ transition. We have also computed the average height of the two consecutive transitions, $P_{\mathrm{LZ},1}$ and $P_{\mathrm{LZ}}(\infty)$, as a function of the phase $\theta$ of the initial
Schr\"odinger-cat state. From the numerical results, it is found that the $\theta$ dependence of $P_{\mathrm{LZ},1}$ and $P_{\mathrm{LZ}}(\infty)$ can be put in the same functional
form but with different fitting parameters.

With sinusoidal driving and an initial photon YS state, the dynamics of the LZSM interferometer is found to be rather complex.
Upon adjustment of the amplitude of the sinusoidal driving, the dynamics of $P_{\mathrm{LZ}}(t)$ is found to change dramatically.
The dynamical behavior of $P_{\mathrm{LZ}}(t)$ can be understood with the help of an energy-level diagram and detailed population dynamics in coupled qubit-photon states.
In addition, we have studied the dependence of LZSM dynamics on the average photon number $|\alpha|^{2}$, and it is revealed that a larger
$|\alpha|^{2}$ makes the qubit more likely stay in the initial up state $\left|\uparrow\right\rangle$ after two consecutive avoided crossings are swept through, because the gap increases with the increasing photon number $n$ of the Fock state, and a larger $|\alpha|^2$ gets higher Fock states involved.

The Schr\"odinger-cat states are important vehicles for implementing quantum computation and quantum error correction process \cite{Lidal_2020,Bonifacio_2020}, and there have been
proposals on using LZSM interferometry to generate Schr\"odinger-cat states \cite{Lidal_2020}.
The process of generating Sch\"odinger-cat states necessarily involves a multitude of Fock states, all of which contribute to the transition probability $P_{\mathrm{LZ}}(t)$. If a
photon state is initially in a Fock state $\left | n \right\rangle$, then the LZ dynamics can be illuminated by the energy diagram or the photon dynamics. As Schr\"odinger-cat
states are superpositions of Fock states with different weights, at any given time, it is not possible to know the contribution of a specific state from an energy diagram alone.

Detailed transition pathways revealed in this work provide insights into quantum state control and monitoring. For instance, the transition pathways present a fundamental physical framework to understand the energy relaxation channels of photoinduced excitation in molecular systems. Understanding the dependence of the qubit states and the transition pathways on the driving field parameters is also helpful for the control of quantum states in quantum information and quantum computing, where photon statistics is often used to readout the qubit states and to identify whether the initial photon field is well prepared.
Detailed analysis of photon dynamics, as demonstrated in this work, help illuminate the contributions from individual states and unveil the intricacies of the LZ transition after the photon state is initiated by a superposition of Fock states.

Qubits usually work in a low temperature environment which can be modeled by a phonon bath with multiple modes. To study the effect of the phonon bath on the qubit dynamics, approaches such as exact diagonalization are incapable to obtain accurate results due to the huge Hilbert space spanned by the multiple phonon modes. The multi-$\mathrm{D}_{2}$ {\it Ansatz} combined with time-dependent variation is an efficient tool to study dephasing and dispersion caused by a low temperature environment with wide-ranging applications in QED devices.

\begin{acknowledgement}
The authors would like to thank D. M.~Jia and Z.~Sun for useful discussion. Support of the Singapore Ministry of Education Academic Research Fund (Grant Nos. 2018-T1-002-175 and
2020-T1-002-075) is gratefully acknowledged.
The work is also supported in part by the Project of Inner Mongolia University of Science \& Technology (2017QDL-B14) and the Natural Science Foundation of Inner Mongolia (2019MS01013).
\end{acknowledgement}

\begin{suppinfo}
  \begin{itemize}
  \item The detailed derivations of the Lagrangian and equations of motion of the multi-D$_{2}$ {\it Ansatz} are given in the Supporting Information.
  \item The convergence test for different initial photon states and the comparison between the multi-D$_{2}$ {\it Ansatz} and the exact diagonalization are shown in the Supporting Information.
    \item The relation between the energy gaps in the vicinity of the avoided crossings and the shorter periods of $P_{n,\,\downarrow}(t)$ ($n>0$) are unveiled in the
        Supporting Information.
  \end{itemize}
\end{suppinfo}
\section{Data Availability}
The data that support the findings of this study are available from the corresponding author upon reasonable request.

%\newpage
\appendix

\beginsupplement
\section{The time dependent variational approach with Davydov states}
\label{Equations of Motion}

Rather than the single mode Hamiltonian in Eq.~(7), we adopt a general Hamiltonian with a multi-mode photon bath
\begin{eqnarray}
    \label{multi_mode_sin_H}
    \hat{H} &=& \frac{\varepsilon(t)}{2} \sigma_{z} + \frac{\Delta}{2} \sigma_{x} +  \sum_{q}^{N} \omega_{q}b_{q}^{\dagger}b_{q} \nonumber\\
    &+&  \sum_{q}^{N} \frac{\gamma_{q}}{2}(\cos\theta_{q} \sigma_{z} + \sin\theta_{q}
  \sigma_{x}) (b_{q}^{\dagger}+b_{q}),\nonumber\\
\end{eqnarray}
where $\omega_{q}$ is the frequency of the $q$th mode of the bath with creation
  (annihilation) operator $b_{q}^{\dagger}$($b_{q}$). $\gamma_{q}$ and $\theta_{q}$ are the qubit-oscillator coupling and the interaction angle, respectively. The multi-mode
  Hamiltonian (\ref{multi_mode_sin_H}) reduces to single-mode Hamiltonian~(7) if $N$ is set to 1.

    The multi-$\mathrm{D}_{2}$ trail state is
  \begin{eqnarray}
    \label{D2_state_mmode}
    |D_{2}^{M}(t)\rangle &=& \sum_{i=1}^{M} \left[A_{i} (t)|\uparrow \rangle \exp(\sum_{q}^{N} f_{iq}(t)b_{q}^{\dagger} - \mathrm{H.c.}) |0\rangle\right] \nonumber \\
    &+& \sum_{i=1}^{M} \left[ B_{i}(t) |\downarrow \rangle \exp(\sum_{q}^{N} f_{iq}(t)b_{q}^{\dagger} - \mathrm{H.c.}) |0\rangle\right], \nonumber \\
  \end{eqnarray}
  in which $A_{i}$ and $B_{i}$ are time-dependent variational parameters for the amplitudes of the up ($\left|\uparrow\right\rangle$) and down
  ($\left|\downarrow\right\rangle$) states, respectively. $f_{iq} (t)$ are the bosonic displacements, where $i$ and $q$ label the $i$th coherent superposition state and
  $q$th effective bath mode, respectively. The trail state (\ref{D2_state_mmode}) reduces to the one mode case in Eq.~(9), if one sets $N=1$.
  If one sets the multiplicity $M=1$, the multi-$\mathrm{D}_{2}$ {\it Ansatz} reduces to the single $\mathrm{D}_{2}$ {\it Ansatz}.
  The single $\mathrm{D}_{2}$ trial state can be seen as a simple direct product of an electronic and a nuclear wave function and can only describe the system state
  in an Ehrenfest approximation. If one let the variational parameters for the nuclear wave function also depend on the electronic part, we get the $\mathrm{D}_{1}$ version trail
  state.
  \begin{eqnarray}
    \label{D1_state_mmode}
    && |D_{2}^{M}(t)\rangle \nonumber\\
    &=& \sum_{i=1}^{M} \left[A_{i} (t)|\uparrow \rangle \exp(\sum_{q}^{N} f_{iq}(t)b_{q}^{\dagger} - \mathrm{H.c.}) |0\rangle\right] \nonumber \\
                         &+& \sum_{i=1}^{M} \left[ B_{i}(t) |\downarrow \rangle \exp(\sum_{q}^{N} g_{iq}(t)b_{q}^{\dagger} - \mathrm{H.c.}) |0\rangle\right], \nonumber \\
  \end{eqnarray}
  in which $f_{iq}$ and $g_{iq}$ are the bosonic displacements for the $\left|\uparrow\right\rangle$ and $\left|\downarrow\right\rangle$ states.
  The multi-$\mathrm{D}_{1}$ {\it Ansatz} of multiplicity $M$ can be viewed as a special case of the multi-$\mathrm{D}_{2}$ {\it Ansatz} of multiplicity 2M.
  If one sets $A_{k}=0$ for even $k$ and $B_{k}=0$ for odd $k$, the multi-$\mathrm{D}_{1}$ {\it Ansatz} becomes the multi-$\mathrm{D}_{2}$ {\it Ansatz} (More details can be
    found in Ref.~45.).

 These multiple Davydov {\it Ans\"atze} in principle allow for an exact solution to the Schr\"odinger equation in the limit of large multiplicities. The
  numerical accuracy and efficiency of the multiple Davydov {\it Ans\"atze} have been extensively verified in a large variety of many-body systems. Although both the
  multi-$\mathrm{D}_{1}$ and the multi-$\mathrm{D}_{2}$ {\it Ans\"atze} are numerically exact with a sufficiently large multiplicity, a suitable version for a particular problem has to be
  carefully chosen depending on the Hamiltonian constructs and parameter configurations. From our extensive studies, we have found that the multi-$\mathrm{D}_{1}$ {\it Ansatz}
  exhibits excellent performance for problems with diagonal system-bath coupling only, while the multi-$\mathrm{D}_{2}$ {\it Ansatz} is more suitable for tasks with off-diagonal
  system-bath coupling, despite that the multi-$\mathrm{D}_{2}$ {\it Ansatz} may have less variational parameters than its $\mathrm{D}_{1}$ counterpart for comparable
  multiplicity. For the Hamiltonian (7), the off-diagonal qubit-photon coupling are employed. Therefore, we adopted the multi-$\mathrm{D}_{2}$ {\it Ansatz} to
  describe the time-dependent state of the entire system.

In order to apply the time-dependent variational principle to explore the dynamics from the Hamiltonian Eq.~(\ref{D2_state_mmode}), we first need to
  calculate the Lagrangian $L$ in Eq.~(2).
\begin{eqnarray}
\label{Lagrangian_detail}
&&L=\frac{i}{2}\sum_{i,j}\left(A_{j}^{*}\dot{A}_{i}-\dot{A}_{j}^{\ast}A_{i}+B_{j}^{\ast}\dot{B}_{i}-\dot{B}_{j}^{\ast}B_{i}\right)S_{ji}\nonumber\\ &&+\frac{i}{2}\sum_{i,j}\left(A_{j}^{\ast}A_{i}+B_{j}^{\ast}B_{i}\right)\sum_{q}[\frac{\dot{f}_{jq}^{\ast}f_{jq}+f_{jq}^{\ast}\dot{f}_{jq}}{2}\nonumber\\
&&-\frac{\dot{f}_{iq}f_{iq}^{\ast}+f_{iq}\dot{f}_{iq}^{\ast}}{2}+f_{jq}^{\ast}\dot{f}_{iq}-f_{iq}\dot{f}_{jq}^{\ast}]S_{ji}\nonumber\\
&&-\left\langle D_{2}^{M}\left(t\right)\right|\hat{H}\left|D_{2}^{M}\left(t\right)\right\rangle,
\end{eqnarray}
where the Debye-Waller factor is $S_{ji}=\exp{\sum_{q}\left\{-\left(\left|f_{jq}\right|^{2}+\left|f_{iq}\right|^{2}\right)/2+f_{jq}^{\ast}f_{iq}\right\}}$,
and the last term in Eq.~(\ref{Lagrangian_detail}) can be obtained as
\begin{eqnarray}
&&\left\langle D_{2}^{M}\left(t\right)\right|\hat{H}\left|D_{2}^{M}\left(t\right)\right\rangle\nonumber\\
  &&=\frac{vt}{2}\sum_{i,j}\left(A_{j}^{\ast}A_{i}-B_{j}^{\ast}B_{i}\right)S_{ji} \nonumber\\
  &&+\frac{\Delta}{2}\sum_{i,j}\left(A_{j}^{\ast}B_{i}+B_{j}^{\ast}A_{i}\right)S_{ji}\nonumber\\
&&+\sum_{i,j}\left(A_{j}^{\ast}A_{i}+B_{j}^{\ast}B_{i}\right)\sum_{q}\omega_{q}f_{jq}^{\ast}f_{iq}S_{ji}\nonumber\\
&&+\frac{1}{2}\sum_{i,j}\left(A_{j}^{\ast}A_{i}-B_{j}^{\ast}B_{i}\right)\sum_{q}\gamma_{q}\cos\theta_{q}\left(f_{iq}+f_{jq}^{\ast}\right)S_{ji}\nonumber\\
&&+\frac{1}{2}\sum_{i,j}\left(A_{j}^{\ast}B_{i}+B_{j}^{\ast}A_{i}\right)\sum_{q}\gamma_{q}\sin\theta_{q}\left(f_{iq}+f_{jq}^{\ast}\right)S_{ji}.\nonumber\\
\end{eqnarray}
The time-dependent variational principle results in equations of motion for $A_i$ and $B_i$,
\begin{eqnarray}
&&-i\sum_{i}\dot{A}{}_{i}S_{ki}\nonumber \\ &&-\frac{i}{2}\sum_{i}A_{i}\sum_{q}\left[-\left(\dot{f}_{iq}f_{iq}^{\ast}+f_{iq}\dot{f}_{iq}^{\ast}\right)+2f_{kq}^{\ast}\dot{f}_{iq}\right]S_{ki}\nonumber \\
  &&=-\frac{vt}{2}\sum_{i}A_{i}S_{ki}-\frac{\Delta}{2}\sum_{i}B_{i}S_{ki} \nonumber \\
  &&-\sum_{i}A_{i}\sum_{q}\omega_{q}f_{kq}^{\ast}f_{iq}S_{ki}\nonumber \\
&&-\frac{1}{2}\sum_{i}A_{i}\sum_{q}\gamma_{q}\cos\theta_{q}\left(f_{iq}+f_{kq}^{\ast}\right)S_{ki}\nonumber\\
&&-\frac{1}{2}\sum_{i}B_{i}\sum_{q}\gamma_{q}\sin\theta_{q}\left(f_{iq}+f_{kq}^{\ast}\right)S_{ki},
\end{eqnarray}
and
\begin{eqnarray}
&&-i\sum_{i}\dot{B}_{i}S_{ki}\nonumber \\ &&-\frac{i}{2}\sum_{i}B_{i}\sum_{q}\left[-\left(\dot{f}_{iq}f_{iq}^{\ast}+f_{iq}\dot{f}_{iq}^{\ast}\right)+2f_{kq}^{\ast}\dot{f}_{iq}\right]S_{ki}\nonumber \\
  &&=+\frac{vt}{2}\sum_{i}B_{i}S_{ki}-\frac{\Delta}{2}\sum_{i}A_{i}S_{ki} \nonumber \\
  &&-\sum_{i}B_{i}\sum_{q}\omega_{q}f_{kq}^{\ast}f_{iq}S_{ki}\nonumber \\
  &&+\frac{1}{2}\sum_{i}B_{i}\sum_{q}\gamma_{q}\cos\theta_{q}\left(f_{iq}+f_{kq}^{\ast}\right)S_{ki}\nonumber\\
&&-\frac{1}{2}\sum_{i}A_{i}\sum_{q}\gamma_{q}\sin\theta_{q}\left(f_{iq}+f_{kq}^{\ast}\right)S_{ki}.
\end{eqnarray}

The equations of motion for $f_{iq}$ are
\begin{eqnarray}
  &&-i\sum_{i}\left[\left(A_{k}^{\ast}\dot{A}_{i}+B_{k}^{\ast}\dot{B}_{i}\right)f_{iq} \right. \nonumber\\
  &&\left . -\left(A_{k}^{\ast}A_{i}+B_{k}^{\ast}B_{i}\right)\dot{f}_{iq}\right]S_{ki}\nonumber\\ &&-\frac{i}{2}\sum_{i}\left(A_{k}^{\ast}A_{i}+B_{k}^{\ast}B_{i}\right)f_{iq}S_{ki}\nonumber\\
&&\times\sum_{p}\left(2f_{kp}^{\ast}\dot{f}_{ip}-\dot{f}_{ip}f_{ip}^{\ast}-f_{ip}\dot{f}_{ip}^{\ast}\right)\nonumber\\
&&=-\frac{vt}{2}\sum_{i}\left(A_{k}^{\ast}A_{i}-B_{k}^{\ast}B_{i}\right)f_{iq}S_{ki}\nonumber\\
&&-\frac{\Delta}{2}\sum_{i}\left(A_{k}^{\ast}B_{i}+B_{k}^{\ast}A_{i}\right)f_{iq}S_{ki}\nonumber\\
  &&-\sum_{i}\left(A_{k}^{\ast}A_{i}+B_{k}^{\ast}B_{i}\right) \nonumber\\
  &&\left(\omega_{q}+\sum\omega_{p}f_{kp}^{\ast}f_{ip}\right)f_{iq}S_{ki}\nonumber\\
  &&-\frac{1}{2}\sum_{i}\left(A_{k}^{\ast}A_{i}-B_{k}^{\ast}B_{i}\right)\gamma_{q}\cos\theta_{q}S_{ki}\nonumber\\
  &&-\frac{1}{2}\sum_{i}\left(A_{k}^{\ast}A_{i}-B_{k}^{\ast}B_{i}\right) \nonumber\\
  &&f_{iq}\sum_{p}\gamma_{p}\cos\theta_{p}\left(f_{ip}+f_{kp}^{\ast}\right)S_{ki}\nonumber\\ &&-\frac{1}{2}\sum_{i}\left(A_{k}^{\ast}B_{i}+B_{k}^{\ast}A_{i}\right)\gamma_{q}\sin\theta_{q}S_{ki}\nonumber\\
  &&-\frac{1}{2}\sum_{i}\left(A_{k}^{\ast}B_{i}+B_{k}^{\ast}A_{i}\right) \nonumber\\
  &&f_{iq}\sum_{p}\gamma_{p}\sin\theta_{p}\left(f_{ip}+f_{kp}^{\ast}\right)S_{ki}.\nonumber\\
\end{eqnarray}
It should be noted that the main results of this work are calculated from the above equations of motion. The equations of motion are solved
numerically by means of the fourth-order Runge-Kutta method.

\section{Convergence test of Landau-Zener dynamics for a qubit coupled to a photon mode}
\label{convergence test}

\begin{figure}
  \centering
  \includegraphics[scale=0.3]{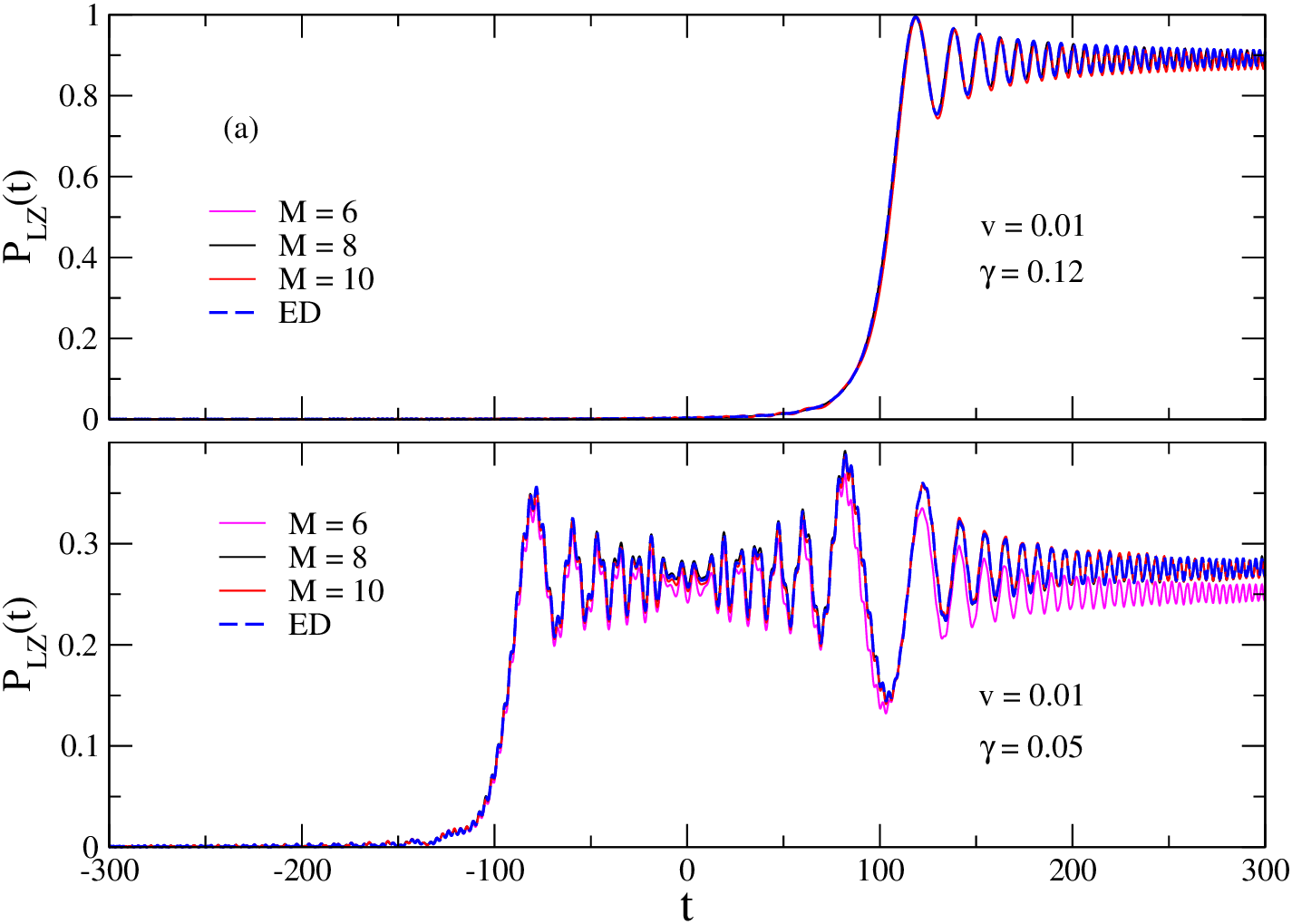}
  \caption{{Time evolution of transition probability calculated by the multi-$\textrm{D}_{2}$ {\it Ansatz} and exact diagonalization. The initial qubit is set to $\left|\uparrow\right\rangle$.
      The multiplicity is set to $M=6$, 8 and 10. Results from the exact diagonalization are plotted with dashed blue curves.
      (a) Initialize the photon bath with the vacuum state $\left| 0 \right\rangle_{\mathrm{ph}}$.
      The coupling strength $\gamma = 0.12$, the speed $v=0.01$. (b) Initialize the photon bath with the YS state in Eq.~(16).
    The coupling strength $\gamma = 0.05$, the speed $v=0.01$. } }
    \label{figS1}
\end{figure}

We perform convergence tests on our algorithm based on the multiple Davydov $\mathrm{D}_{2}$ trial states, and estimate the approximate multiplicity values needed
for our simulations to converge.

First, convergence tests are performed for a single oscillator case in Hamiltonian (8), in order to explore the convergent parameters to obtain accurate results of
the LZ transition probability $P_{\mathrm{LZ}}(t)$.
%The coupling strength $\gamma = 0.5 \sqrt{v/\hbar}$ and frequency $\omega=0.1\sqrt{v/\hbar}$ are employed.

In Fig.~\ref{figS1}(a), a convergence test is done with a vacuum initial state for the photon cavity. The coupling strength $\gamma = 0.12$ and speed $v=0.01$ are
employed. Multiplicities $M$ from {$6$ to $10$} are adopted in the calculations. As shown in Fig.~\ref{figS1}, convergence is achieved for $M=6$ for the vacuum initial photon
state. The steady-state LZ transition probability is found to be 0.883, which agrees with the analytical prediction 0.896 from
\begin{equation}
  P_{\mathrm{LZ}}(\infty) = e^{-\pi \gamma^{2} / 2v}.
\end{equation}

In Fig.~\ref{figS1}(b), a superposition initial photon state in Eq.~(16) with the phase $\theta = \pi/2$, i.e., the YS state, is used to test the convergence. The
coupling strength $\gamma = 0.05$ and speed $v=0.01$ are adopted in the computation. As shown in Fig.~\ref{figS1}(b), multiplicities $M$ from $6$ to $10$ are used
  in the calculations. In Fig.~\ref{figS1}(b), curves of  multiplicities {$M=6$ (pink)} visibly deviate those of higher multiplicities, while
  $P_{\mathrm{LZ}}(t)$ calculated with multiplicities from {$M=8$ to $10$} coincide perfectly.

To provide additional confirmation to the accuracy of our multi-D$_2$ results in this work, the computationally expensive method of exact diagonalization (ED) has been employed as check up, and the results are plotted as blue dashed lines in Fig.~\ref{figS1}
  (a) and (b). It is found that the difference between the ED results and those from the multi-$\mathrm{D}_{2}$ {\it Ansatz} is negligibly small.

  Comparing with the vacuum initial case, a larger multiplicity is needed to achieve
  convergence for the case of a superposition initial photon state. The YS state involves much higher photon energy levels. The difference in the convergence multiplicity between
  the vacuum and YS state implies that involvement of higher photon energy levels requires a larger multiplicity to achieve convergence.

\section{Shorter period of the central oscillation in $P_{n,\downarrow} (t)$ for large $n$}
\label{Delta_E_T_relation}

\begin{figure}[htbp]
    \centering
    \includegraphics[scale=0.3]{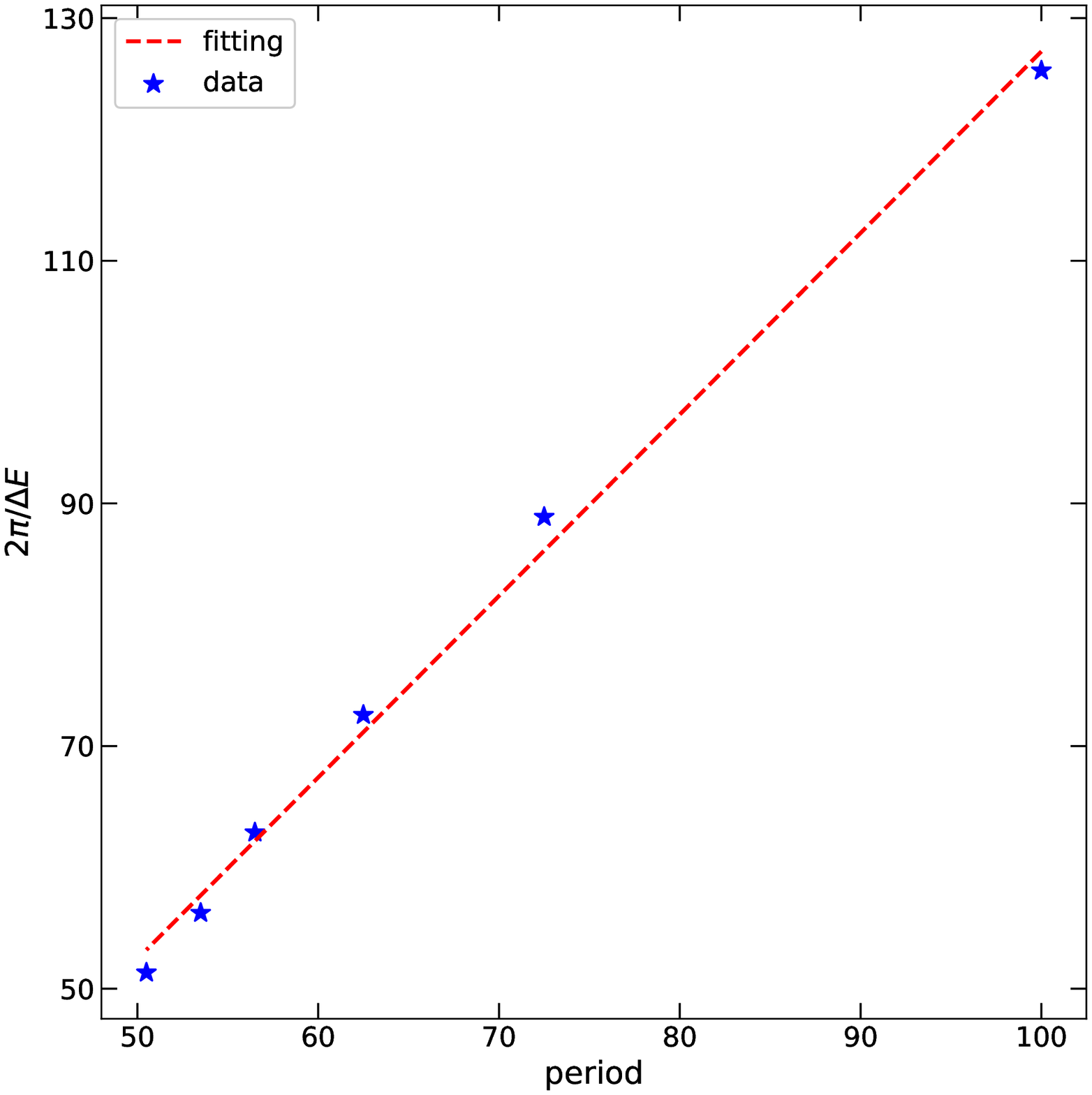}
    \caption{Relation between the energy gaps $\Delta E$ of two nearby adiabatic states at $t=-28$ and the periods of the oscillations around $t=0$ in $P_{n,\downarrow}$. The red dashed line is a linear fitting of the data points.}
    \label{figS2}
\end{figure}

As can be seen from Hamiltonian (6) with an interaction angle $\theta_{\mathrm{c}}=\pi / 2$, due to qubit-photon coupling, energy spaces at avoided crossings between two nearby adiabatic states increase with photon numbers. These energy gaps determine the oscillation periods of $P_{n,\downarrow}$ ($n>0$) in the vicinity of the avoided crossings. To unveil the relations between these two quantities, we measure the energy gaps $\Delta E$ between two nearby adiabatic states at $t=-28$ as well as the oscillation periods around $t=0$ in $P_{n,\downarrow}$ ($n>0$). As shown in Fig.~\ref{figS2}, the inverse of the energy gaps has a perfect linear relation with the oscillation periods. Therefore, shorter-period oscillations are found in populations of states with higher photon numbers at avoided crossings, as depicted in Figs.~7 (e, g, i).

\end{document}